\documentclass[12pt]{article}
\pdfoutput=1
\usepackage{graphicx}
\renewcommand{\baselinestretch}{1.0}

\newcommand{\be}{\begin{equation}}
\newcommand{\ee}{\end{equation}}
\def\aprle{\buildrel < \over {_{\sim}}}
\def\aprge{\buildrel > \over {_{\sim}}}
\newcommand{\bea}{\begin{eqnarray}}
\newcommand{\eea}{\end{eqnarray}}
\begin{document}
\topmargin 0pt
\oddsidemargin=-0.4truecm
\evensidemargin=-0.4truecm
\renewcommand{\thefootnote}{\fnsymbol{footnote}}

\newpage
\setcounter{page}{1}
\begin{titlepage}
\vspace*{-2.0cm}
\begin{flushright}
\vspace*{-0.2cm}

\end{flushright}
\vspace*{0.5cm}

\begin{center}

{\Large \bf Searching for sterile neutrinos in ice}

\vspace{0.5cm}

{Soebur Razzaque$^{1,}$\footnote{Present address: Space Science
Division, U.S. Naval Research Laboratory, 4555 Overlook Ave, SW,
Washington, DC 20375, USA}\footnote{E-mail: srazzaqu@gmu.edu} and
A. Yu. Smirnov$^{2,}$\footnote{E-mail: smirnov@ictp.it}\\
\vspace*{0.2cm} {\em (1) College of Science, George Mason University,
Fairfax, Virginia 22030, USA }\\ {\em (2) The Abdus Salam
International Centre for Theoretical Physics, I-34100 Trieste, Italy }
}

\end{center}

\vskip 1cm

\begin{abstract}
Oscillation interpretation of the results from the LSND, MiniBooNE and
some other experiments requires existence of sterile neutrino with
mass $\sim 1$ eV and mixing with the active neutrinos $|U_{\mu 0}|^2
\sim (0.02 - 0.04)$.  It has been realized some time ago that
existence of such a neutrino affects significantly the fluxes of
atmospheric neutrinos in the TeV range which can be tested by the
IceCube Neutrino Observatory.  In view of the first IceCube data
release we have revisited the oscillations of high energy atmospheric
neutrinos in the presence of one sterile neutrino.  Properties of the
oscillation probabilities are studied in details for various mixing
schemes both analytically and numerically.  The energy spectra and
angular distributions of the $\nu_\mu-$events have been computed for
the simplest $\nu_s-$mass, and $\nu_s - \nu_\mu$ mixing schemes and
confronted with the IceCube data.  An illustrative statistical
analysis of the present data shows that in the $\nu_s-$mass mixing
case the sterile neutrinos with parameters required by LSND/MiniBooNE
can be excluded at about $3\sigma$ level.  The $\nu_s- \nu_\mu$ mixing
scheme, however, can not be ruled out with currently available IceCube
data.
\end{abstract}

\end{titlepage}
\renewcommand{\thefootnote}{\arabic{footnote}}
\setcounter{footnote}{0}
\renewcommand{\baselinestretch}{0.9}

\section{Introduction}

There are several experimental results which could be interpreted as
due to oscillations related to existence of sterile neutrinos with
mass $m \sim 1$ eV and rather large mixing with $\nu_\mu$ or/and
$\nu_e$. This includes the LSND result~\cite{lsnd}, the MiniBooNE
excess of events in neutrino and antineutrino
channels~\cite{miniboone}, the reactor antineutrino
anomaly~\cite{reactor} and the results of the solar calibration
experiments~\cite{calibr} (see \cite{giunti} for recent
interpretation).  Global analysis of the short-baseline oscillation
experiments shows certain consistency of different evidences in the
two sterile neutrinos context~\cite{sglob}.  Furthermore, the analysis
of CMB data indicates an existence of additional radiation in the
Universe~\cite{cmbn} with sterile neutrino being one of the plausible
candidates.  The effective number of neutrino species, $N_{\rm eff}
\sim 4$-5, looks preferable.  The bound on $N_{\rm eff}$ from the Big
Bang Nucleosynthesis (BBN) has been relaxed recently allowing for 1-2
additional neutrinos with the best fit value above 3
species~\cite{bbn}.

At the same time, the recent global cosmological analysis which
includes the CMB data, large scale structure and BBN results, shows
that existence of new neutrino species does not relax significantly
the bound on mass of the sterile neutrino \cite{globc}.  For $\Delta
N_{\rm eff} = 1$ one obtains $m_s < (0.5 - 0.6)$ eV or $\Delta m^2 <
(0.25 - 0.36)$ eV$^2$ which is smaller than the LSND-required value.
 
It has been observed some time ago that mixing of sterile neutrinos
with $m \sim 1$ eV, and therefore $\Delta m^2 \sim 1$ eV$^2$, strongly
affects the atmospheric neutrino fluxes in the energy range 500
GeV-few TeV.  In this energy range the MSW resonance in matter of the
Earth is realized in the $\nu_\mu - \nu_s$ or $\bar{\nu}_\mu -
\bar{\nu}_s$ channel~\cite{nunokawa}.  The resonance enhancement of
oscillations leads to appearance of a dip in the energy spectrum and
to distortion of the angular dependence of tracking
($\nu_\mu-$induced) events. These effects can be studied in the
IceCube detector~\cite{nunokawa}. Later in~\cite{choubey} an extended
study of the oscillation probabilities has been performed in the
presence of one or two sterile neutrinos.  As an experimental test it
has been proposed to measure the ratio of the tracking and cascade
(induced by $\nu_e$) events.

Recently AMANDA \cite{amanda} and IceCube~\cite{icecube} have
published the first high statistics data on the atmospheric neutrinos
in the TeV range.  (See also results from SuperKamiokande
\cite{skmu}).  In this connection we present both analytical and
numerical study of properties of the relevant oscillation
probabilities for different mixing schemes.  We compute the energy
spectra and angular distributions of events in IceCube.  Results of
these computations are confronted with the IceCube data and bounds on
the parameters of sterile neutrinos have been obtained. We show that
observational results substantially depend in the $\nu_s-$ mixing
scheme.

The paper is organized as follows. In Sec.~2 we describe the simplest
mixing scheme for sterile neutrino (the $\nu_s-$mass mixing) for which
dynamics of evolution is reduced to the $2\nu-$evolution.  We obtain
the analytical expressions for the oscillation probabilities and
present results of numerical computations of the probabilities.  In
Sec.~3 we study modifications of the atmospheric neutrino fluxes due
to mixing with sterile neutrinos.  We compute the number of events for
IceCube and confront them with experimental data. In Sec.~4 we perform
an illustrative statistical analysis of the data and obtain bounds on
the mixing of sterile neutrinos depending on the sterile neutrino
mass.  In Sec.~5 the oscillation effects are considered in the $\nu_s
- \nu_\mu$ mixing scheme.  We compute the probabilities and zenith
angle distributions of the $\nu_\mu$ events, and perform the
$\chi^2-$analysis.  In Sec.~6 we study dependence of the oscillation
effects on the mixing scheme in the leading order approximation (valid
at high energies).  Conclusions are given in Sec.~7.  In the Appendix
we present explicit expressions for the probabilities in the constant
density case.

\section{Oscillation probabilities in the $\nu_s-$mass mixing scheme}

We will consider mixing of four flavors\footnote{$\nu_s$ can be
treated as the state with zero flavor.} of neutrinos $(\nu_s,
\nu_e, \nu_\mu, \nu_\tau)$ which mix in four mass eigenstates $\nu_i$,
$i = 0, 1, 2, 3$. We assume the neutrino mass hierarchy: $m_0 \gg m_3,
m_2, m_1$, since the opposite situation: $m_3 \approx m_2 \approx m_1
\gg m_0$, with three active neutrinos in the eV range is strongly
disfavored by the cosmological data.  The mass squared differences
equal
$$
\Delta m^2_{03} \equiv (m_0^2 -  m_3^2) 
\sim (0.5 - 3) ~ {\rm eV}^2, ~~~~  
\Delta m^2_{32} \equiv (m_3^2 -  m_2^2) 
\approx 2.5 \cdot 10^{-3}~ {\rm eV}^2,  
$$
as is required by the LSND/MiniBooNE and fixed by the atmospheric
neutrino results.

As we will show, for high energies ($E > 100$ GeV) the electron
neutrino mixing can be neglected in the first approximation in
consideration of the $\nu_{\mu}-$, $\bar{\nu}_{\mu}-$ oscillations.
Therefore the system is reduced to mixing of the three flavor states
$\nu_f^{T} \equiv (\nu_s, \nu_\tau, \nu_\mu)$ in three mass eigenstates
$\nu_{mass}^{T} \equiv (\nu_0, \nu_3, \nu_2)$ as $\nu_f = U_f
\nu_{mass}$, where $U_f$ is the mixing matrix.  

In this section we will consider the simplest mixing scheme when 
$\nu_s$ mixes in the states $\nu_0$ and $\nu_3$
with masses $m_0$ and $m_3$ only.  In this case
\be
\nu_f =   U_f \nu_{mass} =  
U_{23} U_\alpha  \nu_{mass}.  
\label{mass_scheme}
\ee
Here $U_{23}$ is the usual 2-3 rotation on the angle $\theta_{23} \approx 
45^{\circ}$ and $U_\alpha$ is the rotation of the mass states $\nu_0$
and $\nu_3$ on the angle $\alpha$. Explicitly,
\be
U_f = U_{23} U_\alpha =  
\left(\begin{array}{ccc}
\cos \alpha  &   \sin \alpha  &  0  \\
- \sin \alpha \cos \theta_{23}  & \cos \alpha \cos \theta_{23} 
&  \sin \theta_{23} \\
\sin \alpha \sin \theta_{23} & - \cos \alpha \sin \theta_{23}  
& \cos \theta_{23}
\end{array} \right).
\label{eq:mix-f}
\ee
The sterile neutrino mixing is characterized by a single new mixing
parameter.  In what follows we will refer to (\ref{eq:mix-f}) as to
the $\nu_s-$mass mixing scheme in contrast to the
$\nu_s-$flavor-mixing scheme which will be discussed in Sec.~5.  The
simplest mixing scheme allows us to reduce dynamics of the
$3\nu-$evolution to $2\nu-$evolution exactly.  Other schemes allow to
do this only approximately.

According to (\ref{eq:mix-f}) $\nu_s$ mixes with the state 
\be
\nu_{\tau}^{\prime} \equiv \cos \theta_{23} \nu_\tau - 
\sin \theta_{23} \nu_\mu, 
\label{eq:taupr}
\ee
and there is no mixing of $\nu_s$ with the orthogonal combination: 
\be
\nu_{\mu}^{\prime} \equiv \cos \theta_{23} \nu_\mu  + 
\sin \theta_{23} \nu_\tau. 
\label{eq:mupr}
\ee
Thus, 
$$
\nu_0 = \cos \alpha~\nu_s - \sin \alpha ~\nu_{\tau}^{\prime}, ~~~
\nu_3 =   \cos \alpha ~\nu_{\tau}^{\prime} + \sin \alpha~\nu_s,~~~   
\nu_2 =  \nu_{\mu}^{\prime}. 
$$
In the first approximation at high energies the dominant effect is due
to oscillations driven by the largest mass splitting, $\Delta
m_{03}^2$. Therefore the transitions are described by the flavor
mixing in the $\nu_0$ state.  The corresponding elements of mixing
matrix equal
\be
U_{s0} = \cos \alpha, ~~ U_{\mu 0} = \sin \alpha \sin\theta_{23}, 
~~U_{0\tau} = - \sin \alpha \cos\theta_{23}. 
\label{eq:u-elements}
\ee
The mass squared difference $\Delta m_{32}^2$ gives sub-leading
effects at high energies.  But it produces the leading effects at low
energies ($E < 0.5$ TeV).

Consider evolution of this system in the propagation basis defined as 
\be 
\tilde{\nu}^T \equiv (\nu_s, \nu_{\tau}^{\prime}, \nu_{\mu}^{\prime}).   
\label{eq:propb}
\ee 
It is related to the mass basis as $\tilde{\nu} = U_{\alpha}
\nu_{mass}$, and therefore the evolution equation for $\tilde{\nu}$
reads
\be
i \frac{d \tilde{\nu}}{ dx} = \tilde{H} \tilde{\nu}  
= (U_\alpha H_0^{diag} U_\alpha^T + V)\tilde{\nu}.    
\label{eq:eveq}
\ee 
Here $H_0^{diag} \equiv {\rm diag} (m_0^2, m_3^2, m_2^2)(2E)^{-1}$,
and $V \equiv {\rm diag} (- V_\mu, 0 ,0)$ is the matrix of the
potentials.  In $V$ we have subtracted the matrix $V_\mu {\bf I}$
proportional to the unit matrix ${\bf I}$.  In this way we factor out
the 2-3 mixing from the evolution of neutrino system.  (For earlier
work on evolution of 3 and more neutrino states in matter, selection
of the propagation basis see~\cite{early}).  For neutrinos in the
electrically neutral medium:
$$
V_\mu = V_{\tau} = - \frac{1}{\sqrt{2}} G_F n_N (1 - Y_e) 
= - \frac{1}{\sqrt{2}} G_F n_n,
$$
where $n_N \equiv \rho /m_N$ is the total number density of nucleons,
$n_n$ is the number density of neutrons and $Y_e$ is the number of
electrons per nucleon in the medium.  In the electrically and
isotopically neutral medium $n_n = n_p = n_e$.  Therefore $V_\mu = 0.5
V_e$, where $V_e$ is the difference of potentials for the $\nu_e -
\nu_\mu$ system. For antineutrinos: $\overline{V}_\mu = -
V_\mu$. Explicitly the Hamiltonian is given by
\be
\tilde{H} = 
\left(\begin{array}{ccc}
\frac{\Delta m_{03}^2}{2 E}\cos^2 \alpha - V_{\mu} & 
- \frac{\Delta m_{03}^2}{4 E} \sin 2\alpha & 0  \\
- \frac{\Delta m_{03}^2}{4 E} \sin 2\alpha  &  
- \frac{\Delta m_{03}^2}{2 E} \sin^2 \alpha & 0 \\  
0 & 0 & - \frac{\Delta m_{32}^2}{2 E}
\end{array}
\right).  
\label{eq:halpha1}
\ee
Here again we have subtracted the matrix proportional to the unit
matrix $(m^2_3/2E) {\bf I}$.

The MSW-resonance condition reads
$$
\frac{\Delta m_{03}^2}{2 E} \cos 2\alpha  = V_\mu,  
$$
and since $V_\mu < 0$ the resonance is realized in the antineutrino
channel. The resonance energy $E \sim (2 - 5)~{\rm TeV}~(\Delta
m_{03}^2/ 1 {\rm eV}^2)$, (see the level crossing scheme in
\cite{orlando}).  The state $\nu_{\mu}^{\prime}$ decouples and is not
affected by matter.  It evolves independently as
\be
A_{\mu^{\prime} \mu^\prime} = e^{i\phi_{32}}, ~~~~ 
\phi_{32} =  \frac{\Delta m_{32}^2 x}{2E}. 
\label{eq:mupmup}
\ee

As follows from the form of the Hamiltonian (\ref{eq:halpha1}) the
evolution matrix (matrix of amplitudes) in the propagation basis can
be written as
\be
\tilde{S} =
\left(\begin{array}{ccc}
A_{ss}  & A_{s \tau^{\prime}}  & 0  \\
 A_{\tau^{\prime} s}   &  A_{\tau^{\prime} \tau^{\prime}}  &  0  \\
0 & 0   &  A_{\mu^{\prime} \mu^{\prime}}
\end{array}
\right)
\label{eq:tilds}
\ee
(no $\nu_\mu^{\prime}-$transitions).  From unitarity of $\tilde{S}$ we
have:
\be
|A_{ss}|^2 + |A_{s \tau^{\prime}}|^2 = 1, ~~~~
|A_{\tau^{\prime} s}|^2  +  |A_{\tau^{\prime}  \tau^{\prime}}|^2 = 1, ~~~~
|A_{\mu^{\prime}  \mu^{\prime}}|^2 = 1. 
\label{eq:unitar}
\ee

According to (\ref{eq:taupr}) and (\ref{eq:mupr}) the states of the
propagation basis $\tilde{\nu}$ are related to the flavor states
$\nu_f$ as
\be
\tilde{\nu} = U_{23}^T \nu_f . 
\label{eq:tildeu}
\ee
Therefore the $S$ matrix in the flavor basis $\nu_f$ is  
$$
S = U_{23} \tilde{S} U_{23}^T.  
$$
Using (\ref{eq:tildeu}) and (\ref{eq:tilds}) we obtain 
\be
S =
\left(\begin{array}{ccc}
A_{ss}  &  \cos \theta_{23} A_{s \tau^{\prime}} & 
- \sin \theta_{23} A_{s \tau^{\prime}} \\
\cos \theta_{23} A_{\tau^{\prime} s} &  
\cos^2 \theta_{23} A_{\tau^{\prime} \tau^{\prime}} 
+ \sin^2\theta_{23}A_{\mu^{\prime} \mu^{\prime}} &  
- \sin\theta_{23}\cos\theta_{23}(A_{\tau^{\prime}\tau^{\prime}} 
- A_{\mu^{\prime}\mu^{\prime}})\\
- \sin \theta_{23} A_{\tau^{\prime} s} & 
- \sin\theta_{23} \cos\theta_{23} (A_{\tau^{\prime} \tau^{\prime}} 
- A_{\mu^{\prime} \mu^{\prime}})  &  
\sin^2 \theta_{23} A_{\tau^{\prime} \tau^{\prime}} 
+ \cos^2\theta_{23}A_{\mu^{\prime} \mu^{\prime}}
\end{array}
\right).
\label{eq:tilds1}
\ee
Moduli squared of the elements of this matrix give the corresponding
oscillation probabilities.  In particular, the $\nu_\mu - \nu_\mu$
survival probability, $P_{\mu \mu}$, equals
\be
P_{\mu \mu} = \left|\sin^2 \theta_{23} A_{\tau^{\prime} \tau^{\prime}}
+ \cos^2 \theta_{23} A_{\mu^{\prime} \mu^\prime} \right|^2 , 
\label{eq:probcomp}
\ee
and the other oscillation probabilities with participation of
$\nu_\mu$ are
\bea
P_{\mu s} & = & \sin^2 \theta_{23} |A_{\tau^{\prime} s}|^2 = 
\sin^2 \theta_{23} (1 - |A_{\tau^{\prime} \tau^{\prime}}|^2), 
\nonumber \\
P_{\mu \tau} & = & \sin^2 \theta_{23} \cos^2 \theta_{23}
|A_{\tau^{\prime}\tau^{\prime}} - A_{\mu^{\prime}\mu^{\prime}}|^2. 
\eea
These probabilities satisfy the unitarity condition: $P_{\mu s} +
P_{\mu \tau} + P_{\mu \mu} = 1$. Notice that when
$A_{\tau^{\prime}\tau^{\prime}} = - A_{\mu^{\prime}\mu^{\prime}}$, the
transition probability $P_{\mu \tau} = P_{\mu \tau}^{max} = \sin^2
2\theta_{23}$.  In this case $P_{\mu \mu} = \cos^2 2\theta_{23}$.
Then for maximal 2-3 mixing we have 
$P_{\mu \mu} = 0$ and $P_{\mu \tau} = 1$, and correspondingly, $P_{\mu
s} = 0$.
 
Consider properties of the survival probabilities in the neutrino and
antineutrino channels.  Using (\ref{eq:mupmup}) and
(\ref{eq:probcomp}) we obtain
\bea
P_{\mu \mu} & = & \left|\sin^2 \theta_{23} A_{\tau^{\prime} \tau^{\prime}}
+ \cos^2 \theta_{23}  e^{i\phi_{32}} \right|^2 
\nonumber \\
& = & \sin^4 \theta_{23} P_{\tau^{\prime} \tau^{\prime}}
+ 2 \sin^2 \theta_{23} \cos^2 \theta_{23} 
{\rm Re} \left(e^{- i\phi_{32}} A_{\tau^{\prime} \tau^{\prime}}\right)
+ \cos^4 \theta_{23},  
\label{eq:probmumu} 
\eea
where $P_{\tau^{\prime} \tau^{\prime}} = \left| A_{\tau^{\prime}
\tau^{\prime}} \right|^2$ is the $\nu_{\tau}^{\prime} -
\nu_{\tau}^{\prime}$ survival probability.  For antineutrinos we have
$A_{\tau^{\prime} \tau^{\prime}} \rightarrow \bar{A}_{\tau^{\prime}
\tau^{\prime}} = A_{\tau^{\prime} \tau^{\prime}}(V_\mu \rightarrow -
V_\mu)$.  For $E \aprge 1$ TeV the 2-3 phase is small, $\phi_{32} <
3^{\circ} - 4^{\circ}$, so that in the lowest-order approximation
\be 
P_{\mu \mu} \approx \sin^4 \theta_{23} P_{\tau^{\prime} \tau^{\prime}}
+ 2 \sin^2 \theta_{23} \cos^2 \theta_{23}
{\rm Re} \left(A_{\tau^{\prime} \tau^{\prime}}\right)
+ \cos^4 \theta_{23}. 
\label{eq:probmumu1}
\ee
However, the phase $\phi_{32}$ can not be neglected at low energies,
$E \aprle 0.5$ TeV.  Explicit analytic expression for the amplitude
$A_{\tau^{\prime} \tau^{\prime}}$ is given in the Appendix.  In the
absence of mixing with sterile neutrino one has $\alpha = 0$,
$A_{\tau^{\prime} \tau^{\prime}} = 1$, and consequently
(\ref{eq:probmumu}) is reduced to usual vacuum oscillation probability
due to the 2-3 mixing and 2-3 mass splitting.

\begin{figure}[ht]
\begin{center}
\includegraphics[width=13cm]{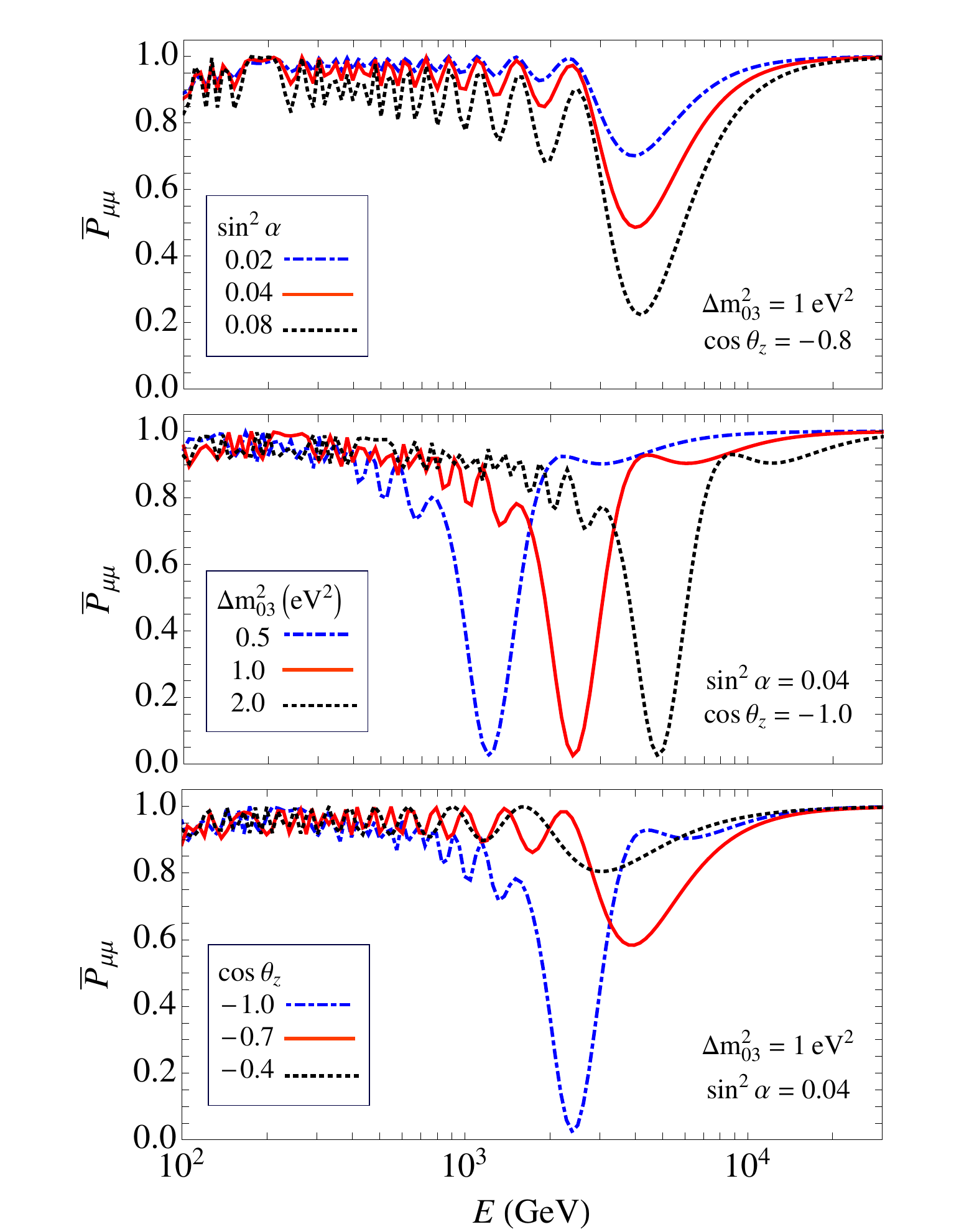}
\caption{The survival probability of the muon antineutrinos (resonance
channel) as function of the neutrino energy for different values of
the zenith angle ($\cos\theta_z$) and oscillation parameters ($\Delta
m_{03}^2$, $\sin^2\alpha$).}
\label{fig:probz1}
\end{center}
\end{figure}

\begin{figure}[ht]
\begin{center}
\includegraphics[width=13cm]{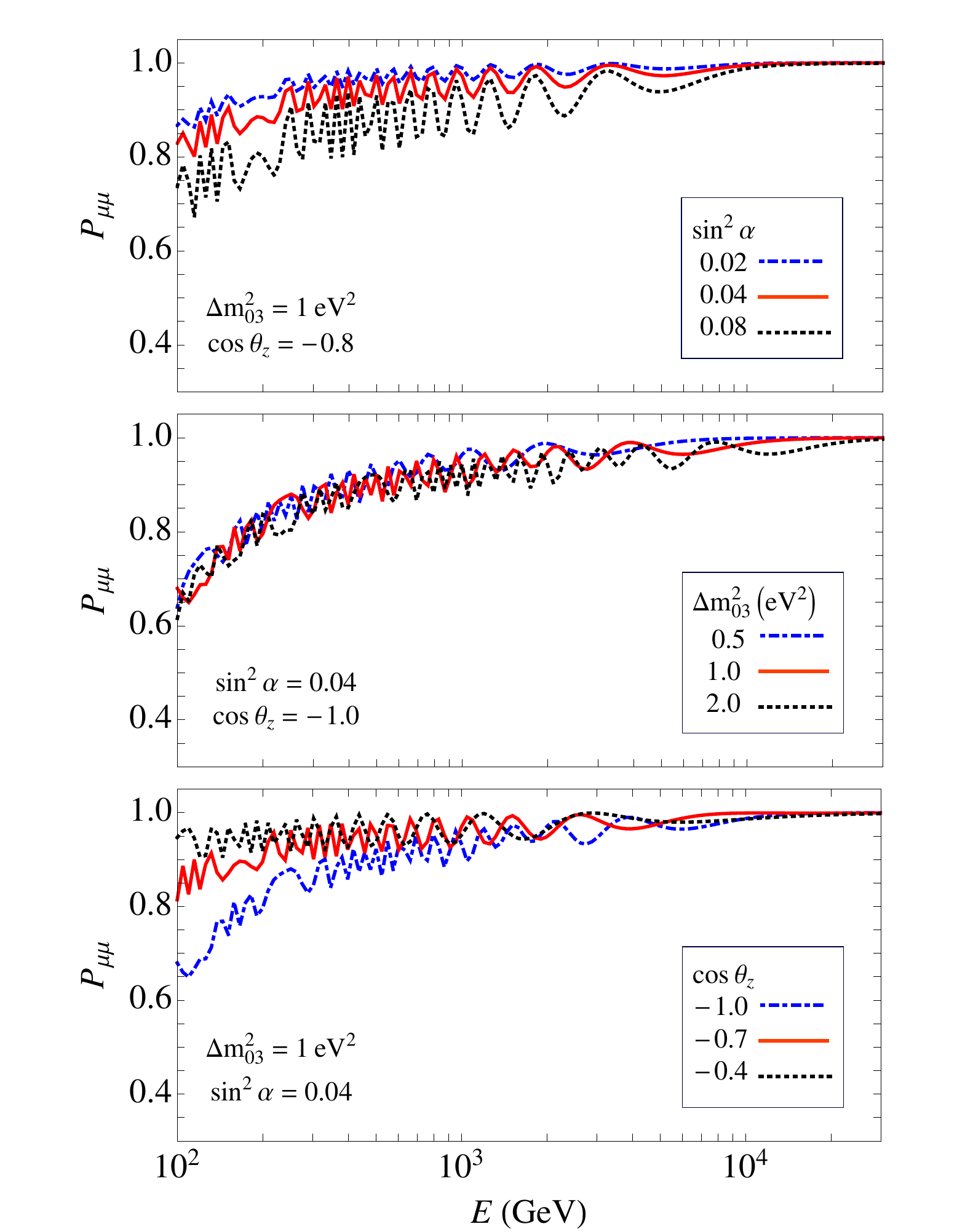}
\caption{The same as in Fig.~\ref{fig:probz1} for muon neutrinos.}
\label{fig:probz2}
\end{center}
\end{figure}

In Fig.~\ref{fig:probz1} we show the probability $\bar{P}_{\mu \mu}$
as a function of neutrino energy for different values of the zenith
angle ($\theta_z$) and the oscillation parameters.  (In our
computations we use the PREM model for the Earth density profile
\cite{prem}.)  The typical energy-dependent feature of $\bar{P}_{\mu
\mu}$ is the resonance dip in the range determined by the resonance
energies in the core and in the mantle.  For $|\cos \theta_z| < 0.82$
there is a single dip at $E \sim E_R \sim 4$ TeV which corresponds to
the MSW resonance in the mantle of the Earth. For $|\cos \theta_z| >
0.82$ (core crossing trajectories) the dependence of the probability
on $E$ is more complicated.  The dip between the resonance energies in
the core and mantle is due to the parametric enhancement of
oscillations, {\em i.e.} due to an interplay between the oscillation
effects in three layers with nearly constant density
(mantle-core-mantle) \cite{param}.  The width of this dip is larger
than the width of the MSW dip in constant density medium.  There is
also the parameteric enhancement of the oscillations at energies above
the resonance energy in the mantle~\cite{param}.

For the $\nu-$ (non-resonance) channel, the peaks are absent (see
Fig.~\ref{fig:probz2}), but another feature related to the matter
effect is realized: enhanced $\mu- \tau$ transition at low ($E < 0.5$
TeV) energies. The survival probability decreases with energy in
contrast to the $\bar{\nu}$ channel where $\bar{P}_{\mu \mu}$
increases with energy.  The reason can be understood from
consideration in the case of constant density (Appendix).  At
energies below 0.5 TeV the oscillations induced by the 2-3 mixing and
mass splitting become important. The dependence of probabilities on
energy is given by the oscillatory curve with low frequency in the
energy scale and the depth $\sin^2 2\theta_{23} \approx 1$ (see
analytic expression in (\ref{eq:mumuconst})).  This curve is modulated
by high frequency oscillations driven by $\Delta m^2_{03}$ with small
depth.  At low energies the phase of the low frequency oscillations is
given (see (\ref{eq:phases12}) in the Appendix) by 
$$
\phi_2 \approx \frac{1}{2}\left(H_{2m}  + 
\frac{\Delta m_{32}^2}{2E} \right) x 
\approx \frac{1}{2}\left(\pm |V_{\mu}| \sin^2 \alpha + 
\frac{\Delta m_{32}^2}{2E} \right) x, 
$$ 
where in last expression the first term is due to the matter effect;
the plus sign corresponds to neutrinos and the minus sign to
antineutrinos (according to (\ref{eq:h2}), $H_{2m} \sim -
V_{\mu} \sin^2 \alpha$). In the energy interval $(0.1 - 0.5)$ TeV the
two contributions are comparable. Thus, the matter effect produces an
opposite change of the phase velocity: increasing the velocity in the
neutrino channel and decreasing it in the antineutrino
channel. Consequently the oscillations due to the 2-3 mass splitting
and 2-3 mixing develop in the $\nu-$ channel at higher energies.
Notice that the phase shift is proportional to $\sin^2 \alpha$, and at
low energies $\alpha_m \approx \alpha$ (see Fig.~\ref{fig:probz2}, the
upper panel).
   
Let us consider more general situation when $\nu_s$
mixes also in the $\nu_2$ state. We introduce an additional rotation
$U_\gamma$ in the $\nu_3$-$\nu_2$ subspace, so that the propagation
basis becomes: $\tilde{\nu} = U_\alpha U_\gamma \nu_{mass}$.
Explicitly the mixing matrix in the propagation basis becomes
\be
\tilde{U}_f = U_\alpha U_\gamma  =
\left(\begin{array}{ccc}
c_\alpha  &   s_\alpha c_\gamma  &    s_\alpha s_\gamma  \\
- s_\alpha  & c_\alpha c_\gamma &  c_\alpha s_\gamma   \\
0 & - s_\gamma  & c_\gamma
\end{array}
\right).
\label{eq:genmix}
\ee
Here $s_\gamma \equiv \sin \gamma$, $c_\gamma \equiv \cos \gamma$, {\it
etc.}.  Now $\nu_s$ mixes in all three mass eigenstates: 
$$
\nu_s = c_\alpha \nu_0 + s_\alpha(c_\gamma \nu_3 + s_\gamma \nu_2). 
$$
The Hamiltonian in the propagation basis equals $\tilde{H}_{\alpha
\gamma} = U_\alpha U_\gamma H^{diag} U_\gamma^T U_\alpha^T$, and it can
be represented as
\be
\tilde{H}_{\alpha \gamma} = \tilde{H} + 
s_\gamma \frac{\Delta m^2_{32}}{2 E}  
\left(\begin{array}{ccc}
- s_\alpha^2 s_\gamma & - s_\alpha c_\alpha s_\gamma  & - s_\alpha c_\gamma \\
... &  - c_\alpha^2 s_\gamma   &  - c_\alpha c_\gamma \\
... & ... &  s_\gamma 
\end{array}
\right),
\label{eq:hamgen}
\ee
where $\tilde{H}$ is the Hamiltonian without $U_\gamma$ rotation
(\ref{eq:halpha1}). The correction is proportional to a small quantity
$s_\gamma \frac{\Delta m^2_{32}}{2 E}$ which produces even smaller
(suppressed by $s_\gamma$) phase than $\phi_{32}$ considered in the
simplest case above. (The matrix in (\ref{eq:hamgen}) is symmetric and
elements denoted by dots equal to the corresponding transponent
elements.)  So, the effects of $\nu_s$ mixing in $\nu_2$ can be
neglected in the first approximation.

The mixing matrix in the flavor basis is given by $U_f = U_{23}
U_\alpha U_\gamma$. The elements of this matrix which describe
oscillations with large mass split $\Delta m^2_{03}$ (dominant at high
energies) are the same as in our simplest mixing case
(\ref{eq:u-elements}).  They do not depend on $\gamma$.

In what follows we present predictions for IceCube in the simplest
$\nu_s-$mass mixing case.  Consideration of the $\nu_s-$flavor mixing
schemes is given in Sec.~5, where we show that, in fact, the
probabilities and observables substantially depend on the mixing
scheme.

\section{Fluxes and numbers of events}

The $\nu_\mu-$flux at the detector equals  
\be  
\Phi_\mu = \Phi_\mu^0 P_{\mu \mu} + \Phi_e^0 P_{e \mu} 
\approx \Phi_\mu^0 P_{\mu \mu},  
\label{eq:fluxmu}
\ee
where $\Phi_\mu^0$ and $\Phi_e^0$ are the original fluxes of $\nu_\mu$
and $\nu_e$ without oscillations.  Similar expression holds for the
antineutrinos.  The effect of $\nu_e \rightarrow \nu_\mu$ oscillations
can be neglected (the last equality in (\ref{eq:fluxmu})).  The reason
is two fold: at high energies $\Phi_\mu^0 \gg \Phi_e^0$, with ratio $r
\equiv \Phi_\mu^0/\Phi_e^0 > 20$ for $E \sim 1$ TeV.  Furthermore, the
transition probability $P_{e \mu} \ll 1$ and $\nu_e$ can be mostly
converted to $\nu_s$.

Let us consider $\nu_e$ oscillations in some details.  At high
energies the mixing of $\nu_e$ and $\nu_\mu^{\prime}$ is strongly
suppressed: $\sin^2 2\theta_{12} (E^R_{12}/E)^2$, where $E^R_{12} \sim
0.1$ GeV is the resonance energy associated to the ``solar'' mass
splitting $\Delta m^2_{21}$.  The $\nu_e - \nu_\tau^{\prime}$ mixing
is absent in the limit $\theta_{13} = 0$, but if non-zero, the 1-3
mixing in matter is also suppressed in the TeV energy range as $\sim
\sin^2 2 \theta_{13} (E_{13}^R/E)^2$, where $E_{13}^R \approx 6$ GeV
is the energy of 1-3 resonance.  Consider the whole $4\nu-$ scheme
with $\nu_e$ admixture, $U_{e0}$, in the state $\nu_0$.  Since for the
$\nu_e$ potential we have $V_e \approx - V_{\mu}$ in the isotopically
neutral medium, the $\nu_e - \nu_s$ level crossing is in the neutrino
channel. The corresponding resonance energy $E_{es}^R \approx E_{\mu
s}^R$.  The depth of $\nu_e - \nu_\mu$ oscillations driven by $\Delta
m_{01}^2$ equals
\be
D_{e\mu}  \approx 4 |U_{e0}^m|^2 |U_{\mu 0}^m|^2,  
\label{eq:depth}
\ee
where $U_{e0}^m$ and $U_{\mu0}^m$ are the mixing parameters in matter.
In vacuum: $D_{e\mu} = \sin^2 2\theta_{\rm LSND} \sim 3 \cdot
10^{-3}$.  The mixing and the depth can be enhanced in resonances.  In
the $\bar{\nu}_\mu - \bar{\nu}_s$ resonance the $\bar{\nu}_\mu-$mixing
is enhanced, $|U_{\mu 0}^m|^2 \sim 1/2$, whereas the $\nu_e-$ mixing
is suppressed: $|U_{e0}^m|^2 \sim |U_{e0}|^2/4$.  As a result,
$D_{e\mu} \approx |U_{e0}|^2/2 \aprle 0.02$.  In the $\nu_e - \nu_s$
resonance, inversely, the $\bar{\nu}_\mu-$mixing is suppressed
$|U_{\mu 0}^m|^2 \sim |U_{\mu 0}|^2/4$, and $\nu_e-$ mixing is
enhanced $|U_{e0}^m|^2 = 1/2$.  So that the depth of oscillations
equals $D_{e\mu} \approx |U_{\mu 0}|^2/2 \aprle 0.02$. Therefore $P_{e
\mu} \aprle 0.02$, and the contribution of the original $\nu_e$ flux
to $\nu_\mu$ flux at a detector, $P_{e \mu} r$, is smaller than
$10^{-3}$.

The rate of $\nu_\mu$ events in a detector such as IceCube is given by
\be
N = \int dE \int d\Omega
\left[\Phi_\mu (E, \theta_z) A_{\rm eff}(E, \theta_z) +
\bar{\Phi}_\mu (E, \theta_z) \bar{A}_{\rm eff}(E, \theta_z)\right],  
\label{eq:murate}
\ee
with the appropriate integrations over the neutrino energy and solid
angle.  Additional contribution to the muon events comes from the
$\nu_\mu \to \nu_\tau$ oscillations, producing a flux $\Phi_\tau
= \Phi^0_\mu P_{\mu\tau}$ at the detector.  The tau lepton from
$\nu_\tau$ interaction has $\approx 18\%$ probability to decay into
muon, which is then recorded as a $\nu_\mu$ event.  The $\nu_\tau$
energy, however, needs to be $\sim 2.5$ times higher than the
$\nu_\mu$ energy to produce muon tracks of the same energy in the
detector.  Notice that in the $\nu_s-$mass mixing scheme $\nu_\tau$'s
appear in the $\nu_\mu$ oscillation dip, but this will lead to
additional events at low energies.  In other mixing schemes
$\nu_\mu$'s are transformed mainly into $\nu_s$'s, and production of
$\nu_\tau$ is suppressed.
 
In (\ref{eq:murate}) $A_{\rm eff}$ and $\bar{A}_{\rm eff}$ are the
effective areas of the detector for $\nu$ and $\bar{\nu}$. They are
given by the effective volume $V_{\rm eff}$ from which the events
(muons) are collected with an efficiency of detection 
$\epsilon_{\rm det}$ as
$$
A_{\rm eff} \sim V_{\rm eff} n_N \sigma_{\nu N} \epsilon_{\rm det}.
$$ 
Here $n_N$ is the number density of nucleons in the surrounding medium
and $\sigma_{\nu N}$ is the neutrino-nucleon charge-current cross
section.  In turn, $V_{\rm eff}$ is determined by the geometry of the
detector and the muon range $R_\mu$: $V_{\rm eff} \propto R_\mu^3$.
The range can be estimated as
$$
R_\mu = \frac{1}{b} \ln 
\frac{a + b E_\nu (1- \langle y \rangle)} {a + b E_{\mu, min}}. 
$$
where $a = 0.24$ GeV~m$^{-1}$, $b = 3.3 \cdot 10^{-4}$~m$^{-1}$,
$\langle y \rangle$ is the mean inelasticity and $E_{\mu, min}$ is the
minimum muon energy for detection.  At low energies $R_\mu \propto
E_\mu$ and at $E \sim 1$ TeV the linear increase of $R_\mu$ changes to
the logarithmic one (see \cite{icecube} for details).  Since, usually
the data are presented using energy bins of equal size in the
$log-$scale, the relevant quantity which determines the number of
events in a given energy bin is $N_{E} = A_{\rm eff} E \Phi_\nu$
(where the $E$ originates from the Jacobian).  The differential
neutrino flux decreases as $\Phi_\nu \propto E^{-3.7}$, and therefore
at low energies $N_E$ increases as $N_{E} \propto E^{1.3}$. It reaches
maximum at $E \sim 0.7$ TeV and then decreases since $V_{\rm eff}$ has
only logarithmic increase.  The median energy interval $E = (0.15 -
2.3)$ TeV is determined by a condition $N_{E} \geq 0.5 N_{E}^{\rm
max}$.  This interval includes the region of dips in the oscillation
probability and therefore IceCube is well optimized to search for
sterile neutrinos with $\Delta m^2 = (0.5 - 2)$ eV$^2$.  The described
dependence of $N_{E}$ on energy allows one to understand various
features of the predicted effects.

The effective area is also given by
$$ 
A_{\rm eff} = A_{\rm det} S_{\rm Earth}(E_\nu, \theta_z) 
P_{\rm int}(E_\nu), 
$$
where $A_{\rm det}$ is the geometrical area of the detector, $S_{\rm
Earth}$ is the survival probability of neutrino passing through the
Earth at a given trajectory and $P_{\rm int}$ is the charged current
neutrino-nucleon interaction probability in the vicinity of the
detector.  The survival probability is given by
$$
S_{\rm Earth} (E_\nu, \theta_z) 
= \exp\left[-N_A \sigma_{\rm tot} (E_\nu) 
\int_0^L \rho(\theta_z, l)dl \right], 
$$
where $L = 2R_{\rm Earth}\cos\theta_z$ is the length of the
trajectory, $\rho(\theta_z, l)$ is the matter density at a distance
$l$ along the trajectory and $\sigma_{\rm tot}$ is the total
neutrino cross-section. For $E\aprle 10$ TeV, $S_{\rm Earth}
\sim 1$. The interaction probability is given by
$$
P_{\rm int}(E_\nu) = N_A \sigma_{\nu N}(E_\nu) \langle R(E_\nu,
E_{\mu, \rm min}) \rangle,
$$
where $ \langle R(E_\nu, E_{\mu, \rm min}) \rangle$ is the average
muon range in the medium and $N_A$ is the Avogadro's number.

\begin{figure}[ht]
\begin{center}
\includegraphics[width=13cm]{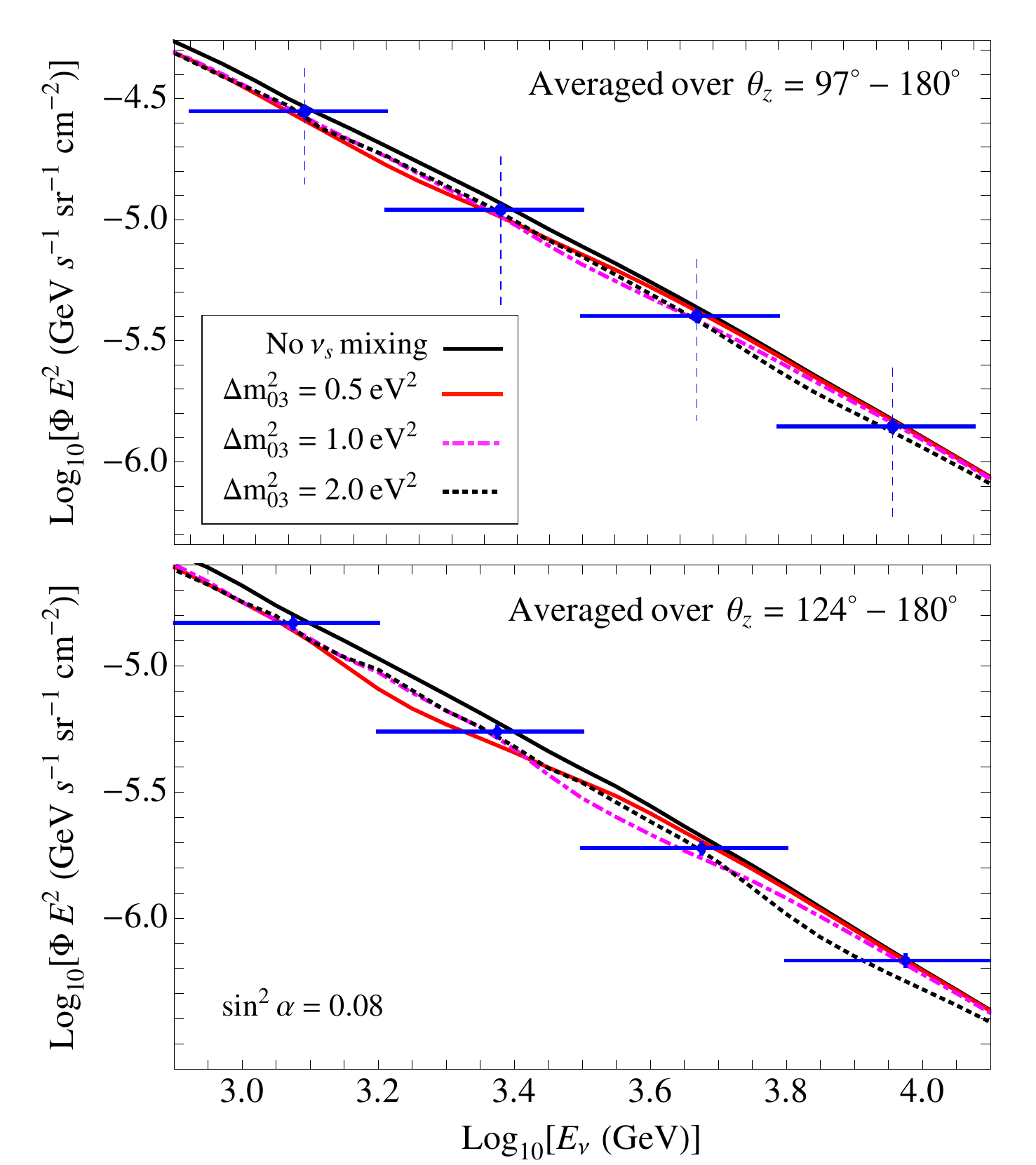}
\caption{The energy spectrum of $\nu_\mu$ integrated over the zenith
angles in the intervals $97^{\circ} - 180^{\circ}$ and $124^{\circ} -
180^{\circ}$ with and without oscillations to sterile neutrinos versus
IceCube result.  We use the $\nu_s-$mass mixing scheme with $\sin^2
\alpha = 0.08$ and $\Delta m_{03} = 1$ eV$^2$.  In the top panel, the
error bars denoted by the dashed lines include both statistical and
systematic errors as reported by Icecube (Table II in \cite{icecube}).
In the lower panels, statistical-only error bars are denoted by solid
vertical lines. For the top panel, the statistical error bars are
about the same size as the points.  }
\label{fig:ice1}
\end{center}
\end{figure}

\begin{figure}[ht]
\begin{center}
\includegraphics[width=13cm]{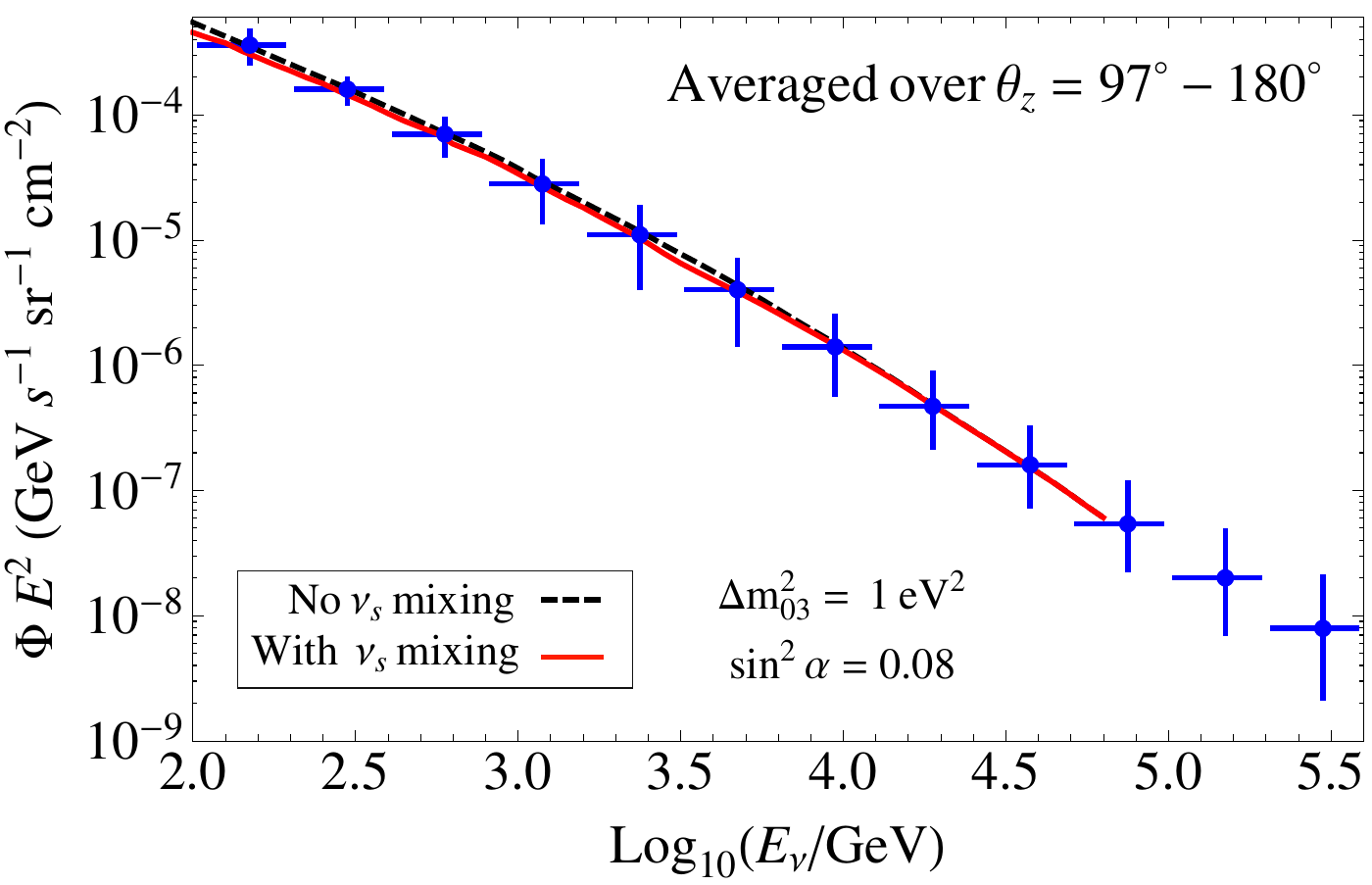}
\caption{The $\nu_\mu$ energy spectrum integrated over the zenith
angle with and without oscillations to sterile neutrinos versus the
IceCube results.  We use $\sin^2 \alpha = 0.08$ and $\Delta m_{03}^2 =
1$ eV$^2$.  }
\label{fig:ice2}
\end{center}
\end{figure}

\begin{figure}[ht]
\begin{center}
\includegraphics[width=13cm]{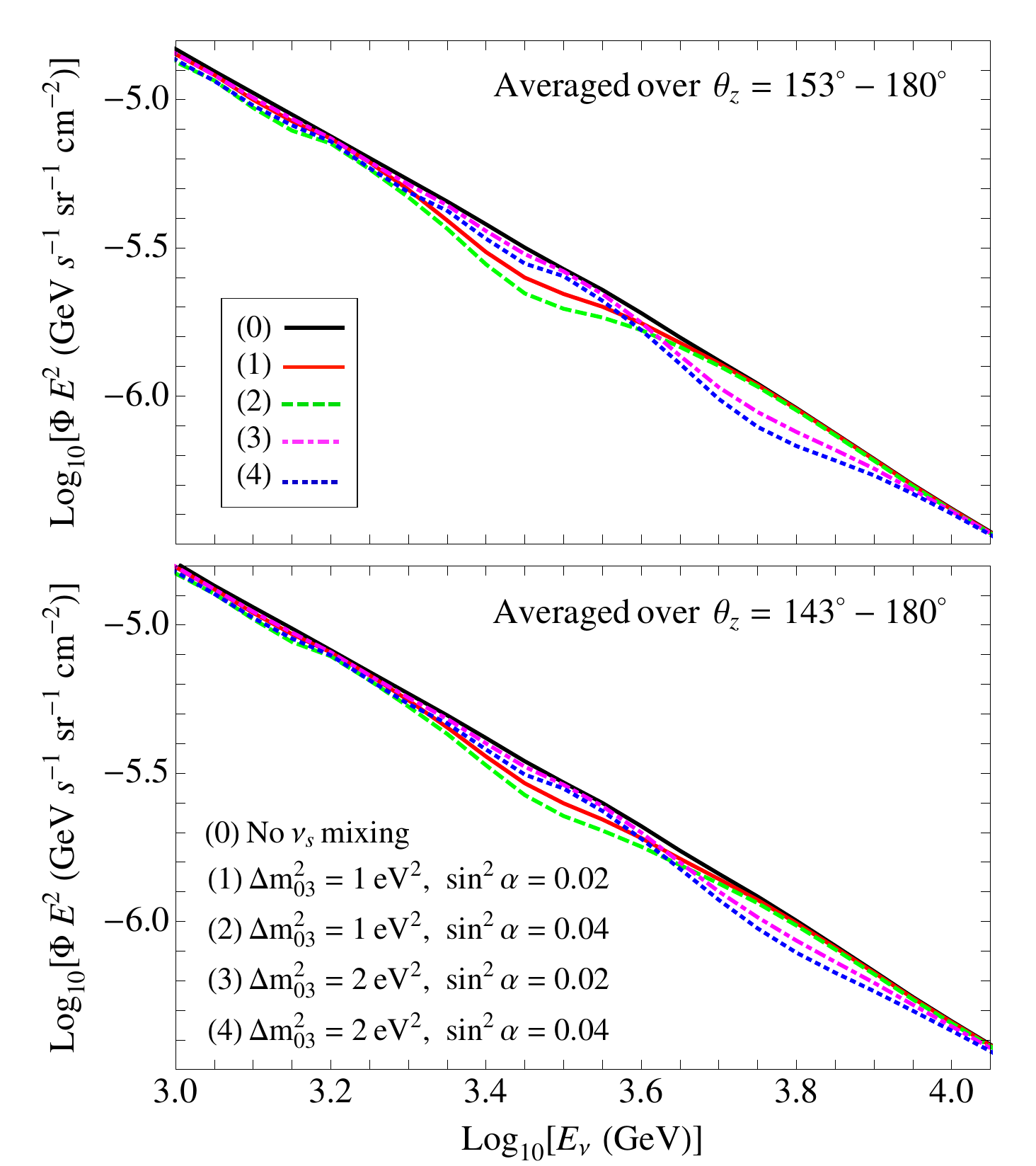}
\caption{The $\nu_\mu$ energy spectrum integrated over different
intervals of zenith angles for different values of $\sin^2 \alpha$ and
$\Delta m_{03}^2$.}
\label{fig:intz}
\end{center}
\end{figure}

In Figs.~\ref{fig:ice1} and \ref{fig:ice2} we show the sum of the
$\nu_\mu$ and $\bar{\nu}_\mu$ energy spectra integrated over the solid
angle for the $\nu_s-$mass mixing scheme.  
An estimation of the size of the oscillation effects is rather
easy: maximal, $\approx 100\%$, effect is for $\bar{\nu}$ in the
resonance range; summation with $\nu$ (whose flux is about 1.4 times
larger) gives $40\%$ effect; averaging over the zenith angle from
$180^{\circ}$ to $90^{\circ}$ produces another factor $\sim 1/2$, and
therefore one arrives at the maximal $\sim 20\%$ suppression in the
dip.  Relative effect increases with narrowing the integration region
around vertical direction (see Fig.~\ref{fig:intz}).  Now the maximal
effect can reach $40\%$ and further enhancement would require
experimental separation of neutrino and antineutrino signals.  With
increase of $\Delta m^2_{03}$ the dip shifts to high energies as $E
\propto \Delta m^2_{03}$. Increase of the size of the dip with $\sin^2
\alpha$ is more complicated.  Suppression effect extends to low
energies due to oscillations in the $\nu-$ channel driven by the 2-3
mixing.

We also compare the predicted neutrino energy spectra in
Figs.~\ref{fig:ice1} and \ref{fig:ice2} with the ``unfolded'' energy
spectra reconstructed by IceCube \cite{icecube}.  Presently, this
comparison can be used for illustration only since reconstruction of
the unfolded spectra implies significant smearing and in general is
not sensitive to the spectral distortion in small energy intervals.
Notice, however, that the size of the dip in the energy scale is
larger than the size of the bin of the reconstructed spectrum. To have
better sensitivity to the distortion one can further decrease the size
of the bin.

According to the Fig.~22 of \cite{icecube} the statistical error in
the relevant energy range is about $3\%$ which is substantially
smaller than the size of the dip.  Continued operation of IceCube in
future will reduce this error further.  Large errors are due to
systematics: mostly due to uncertainties in the total normalization
and tilt of the spectrum.  To a large extent they can be eliminated
when searching for the dip.  Indeed, the systematics has smooth
dependence on energy, the systematic errors in different bins are
strongly correlate.  One can parametrize these uncertainties by a few
parameters and determine them by fitting data.

\begin{figure}[ht]
\begin{center}
\includegraphics[width=11cm]{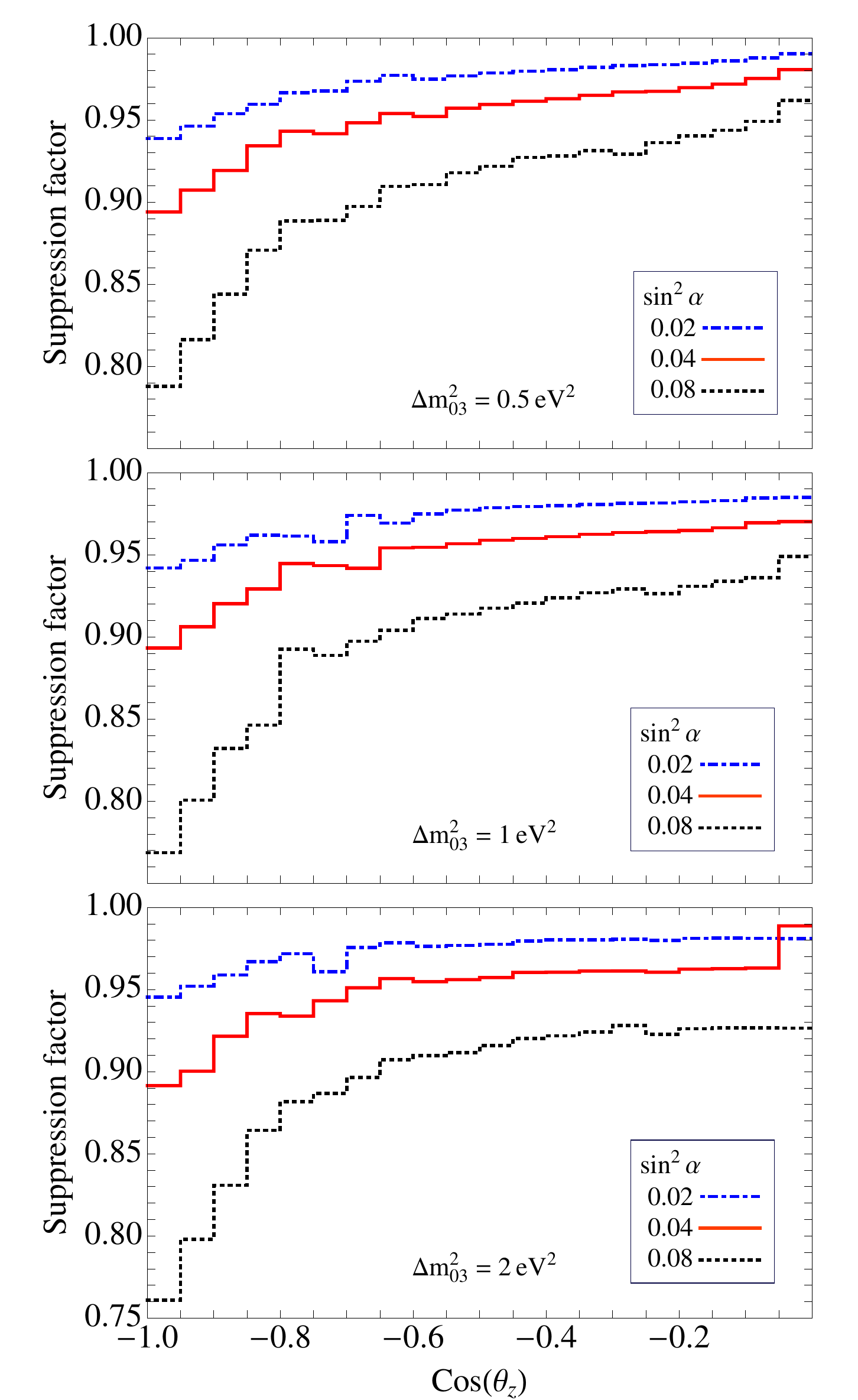}
\caption{The zenith angle dependence of the suppression factor of the
muon events integrated over the energy from $E_{th} =0.1$ TeV.  We use
the $\nu_s-$mass mixing scheme.}
\label{fig:zen01}
\end{center}
\end{figure}

\begin{figure}[ht]
\begin{center}
\includegraphics[width=11cm]{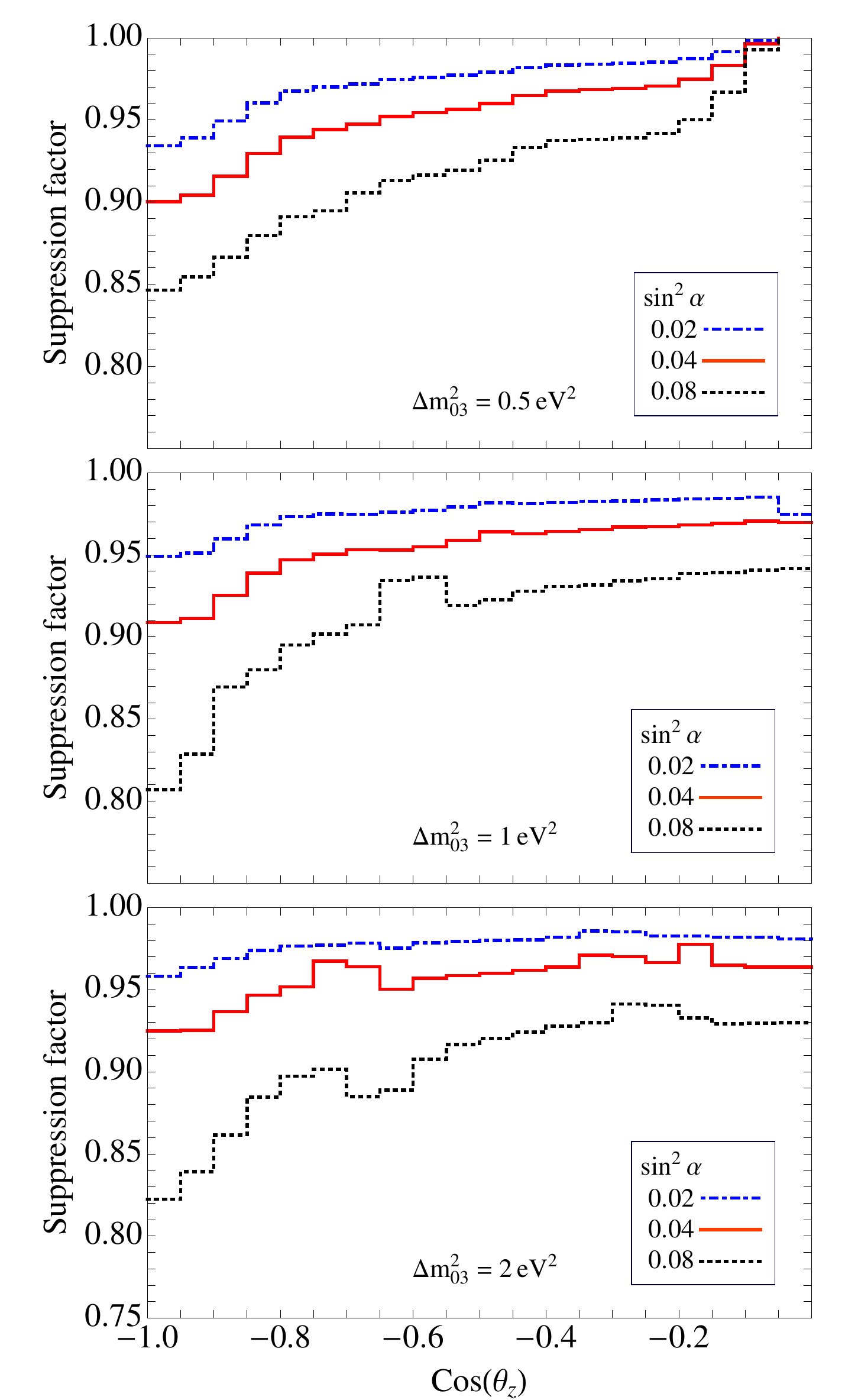}
\caption{The same as in fig. \ref{fig:zen01} for $E_{th} = 1$ TeV.}
\label{fig:zen1}
\end{center}
\end{figure}

The problem of smearing does not exist in the case of the zenith angle
distribution, since muons nearly follow neutrinos, and the zenith
angle resolution is $0.5 - 1^{\circ}$. We compute the number of events
$N_j$ in a given zenith angle bin $\Delta_j \cos \theta_z$ using
(\ref{eq:murate}) and performing integration from the threshold
$E_{th}$:
\be 
N_j = 2\pi \int_{\Delta_j \cos\theta_z} d\cos\theta_z
\int_{E_{th}} dE~ 
\Phi_\nu^0 (E,\theta_z) A_{\rm eff} (E,\theta_z) P_{\mu\mu}(E,\theta_z)
+ {\rm antineutrinos}. 
\label{eq:nj}
\ee 
We then define the suppression factors in the individual bins as  
\be
S_j = \frac{N_j}{N^0_j}, 
\label{eq:indb}
\ee
where $N^0_j$ are the numbers of events without oscillations which
correspond to $P_{\mu\mu} = 1$ in (\ref{eq:nj}).  In
Figs.~\ref{fig:zen01} and \ref{fig:zen1} we show the zenith angle
dependence of the suppression factor for different values of the
mixing parameter $\sin^2 \alpha$ and $\sin^2 \theta_{23} = 1/2$ (this
corresponds to $|U_{\mu 0}|^2 = 0.5 \sin^2 \alpha$) and two different
thresholds $E_{th} = 100$ GeV (Fig.~\ref{fig:zen01}) and $E_{th} = 1$
TeV (Fig.~\ref{fig:zen1}).  Oscillations lead to distortion of the
zenith angle distribution.  For nearly horizontal direction the effect
is mainly due to vacuum oscillations which have enough baseline to
develope if $E \aprle 0.5$ TeV. In this case the averaged oscillation
effect is given by $1 - 2 |U_{\mu 0}|^2 (1 - |U_{\mu 0}|^2) \approx 1
- \sin^2 \alpha$ in agreement with the results of
Figs.~\ref{fig:zen01} and ~\ref{fig:zen1}.  The matter effect
increases with $|\cos \theta_z|$.  According to these figures
substantial differences between the energy-integrated distribution
with and without sterile mixing are expected in the bins near the
vertical direction.  For $\sin^2 \alpha = 0.04$ the effect is about
$20 \%$ and the statistical errors, $3 \%$, are much smaller.  For
other mixing schemes the distortion can be different. In particular,
in the $\nu_s-$flavor mixing scheme maximal suppression is in the bins
$\cos \theta_z = (- 0.9, - 0.8)$ (see Sec.~5).

For vertical directions the evaluation of the suppression (integrated
over the energy) can be done using the survival probabilities of
Figs.~\ref{fig:probz1} and \ref{fig:probz2}. If {\em e.g.} $\sin^2
\alpha = 0.08$, the probabilities averaged over the median energy
interval in the neutrino and antineutrino channels are $P_{\mu \mu} =
0.6$ and $\bar{P}_{\mu \mu} = 0.8$ respectively.  Then averaging the
contributions of the neutrinos and antineutrinos we obtain $S \sim
\langle P \rangle = 0.70 - 0.75$, in agreement with results in
Figs.~\ref{fig:zen01} and \ref{fig:zen1}.  With increase of threshold,
the effect of vacuum oscillations in nearly horizontal directions
becomes smaller. The effect in the $\bar{\nu}$ channel increases,
whereas in the $\nu$ channel it decreases, thus compensating the
overall change.

\begin{figure}[ht]
\begin{center}
\includegraphics[width=11cm]{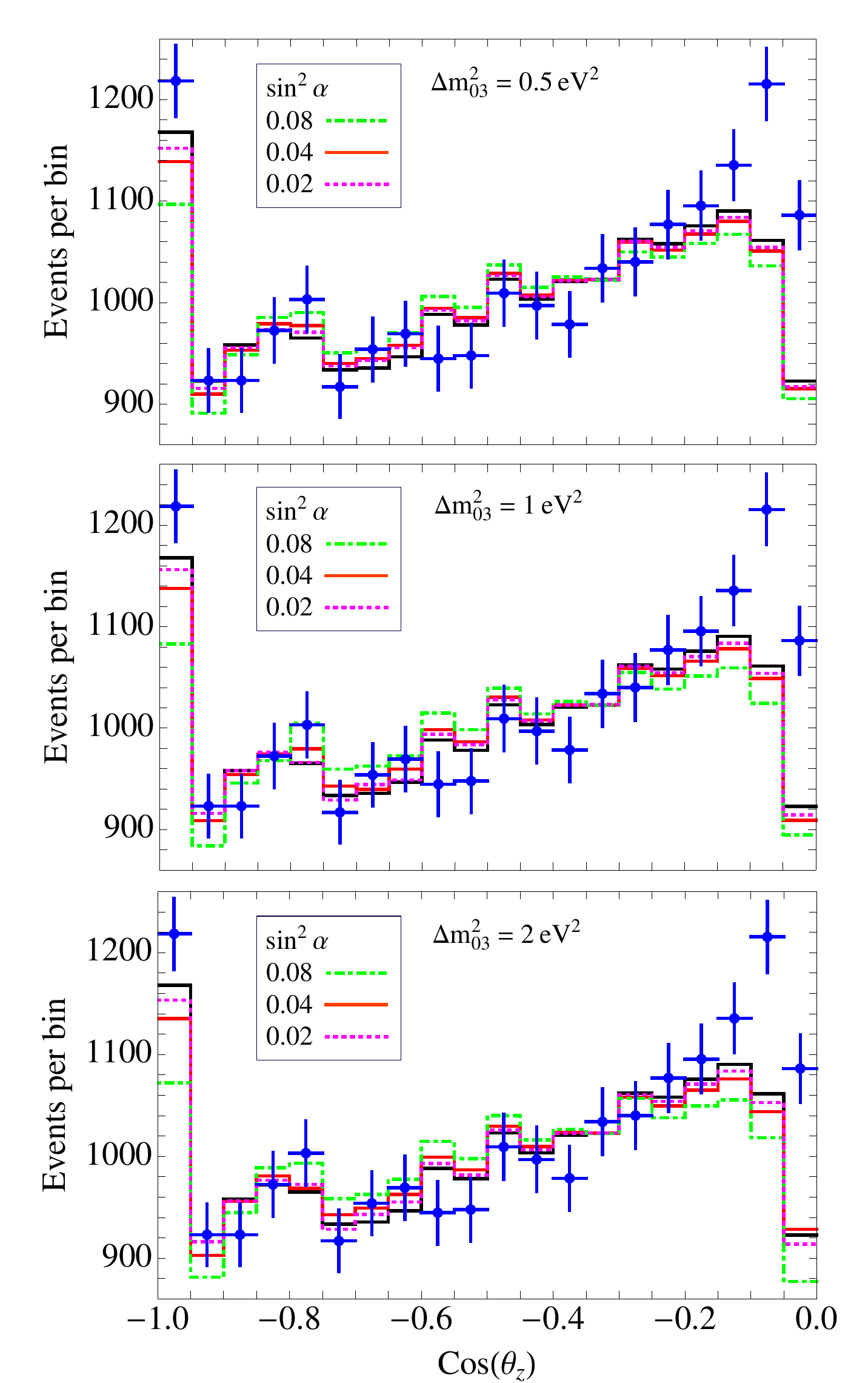}
\caption{The zenith angle distribution of muons from $\nu_\mu$
interactions integrated over the energy with and without oscillations
(solid black histograms) to sterile neutrinos.  We have renormalized
the event distribution according to the best-fit normalization and
tilt parameters from the $\chi^2$ fit (Table 1).  Also shown are the
IceCube results.}
\label{fig:ice8}
\end{center}
\end{figure}

In Fig.~\ref{fig:ice8} we confront the experimental results with 
the predicted zenith angle distributions computed as  
$$
N_j = N^{MC}_j S_j,  
$$
where $N^{MC}_j$ is taken according to the IceCube simulation (see
Fig.~19 from~\cite{icecube}).  We have implemented an overall
normalization and tilt of the distribution, as we discuss below in
(\ref{renormalize}).

\section{Bounds on parameters of sterile neutrinos}

To get an idea of the sensitivity of the currently available IceCube
data to the sterile neutrino mixing we have performed a $\chi^2$ fit
of the IceCube zenith angle distribution.  For a given ``model'' of
mixing characterized by $(\Delta m_{03}^2,~ \sin^2\alpha)$ we compute
the expected number of muon events $N_j^{\rm mod}$ in the zenith angle
bin $j$.  For this we use the IceCube simulation, $N_j^{\rm MC}$
\cite{icecube}:
\begin{equation}
N_j^{\rm mod} (C, \tau; \Delta m_{03}^2, \sin\alpha) = 
C [1 + \tau (\cos\theta_j + 0.5)] N_j^{\rm MC}
S_j(\Delta m_{03}^2, \sin^2\alpha) \,, 
\label{renormalize}
\end{equation}
where $C$ is an overall normalization parameter and $\tau$ is a
zenith angle tilt parameter.  The model without $\nu_s$ mixing is
recovered when $\alpha = 0$.  We compare the expected numbers
$N_j^{\rm mod}$ with data $N_j^{\rm dat}$ and the $\chi^2$ is defined
as
\be
\chi^2 (C, \tau; \Delta m_{03}^2, \sin\alpha) = 
\sum_j 
\frac{\left( N_j^{\rm dat} -  
N_j^{\rm mod} (C, \tau; \Delta m_{03}^2, \sin\alpha)\right)^2 }
{\left( \sigma_j^{\rm dat} \right)^2} \, . 
\ee
The variance $\sigma_j^{\rm dat}$ is calculated by adding in
quadrature the statistical and systematic uncertainties as given by 
IceCube \cite{icecube}.  For our analysis we use the
IceCube data in the range of zenith angles $-1\le \cos\theta_z \le
-0.1$ ({\it i.e.}, bins $j = 1$-18), leaving out the last two near
horizontal bins where the detector response is not well-understood and
contamination of the atmospheric muons is possible.  For fixed values
$(\Delta m_{03}^2, \sin\alpha)$ we minimize the $\chi^2$ varying the
$(C, \tau)$ parameters. The difference
$$
\Delta\chi^2 = \chi^2_{\rm min} (C, \tau; \Delta m_{03}^2,
\sin^2\alpha) - \chi^2_{\rm min} (C, \tau; \Delta m_{03}^2, 0)
$$ 
quantifies the rejection significance of the $\nu_s$ mixing model
with respect to the model without $\nu_s$ mixing.

In Table~1 we show results of our statistical analysis which is
reduced to determination of the minimal $\chi^2$ values of $C$ and
$\tau$ for given $\Delta m_{03}^2$ and $\sin^2\alpha$.  We show
$\chi^2_{\rm min}$ and the best fit values of $C$ and $\tau$ for the
case of statistical errors for the individual bins only (see
Fig.~\ref{fig:ice8}).  Also shown is the fit for the ``null''
hypothesis.  Notice that the $\nu_s$ mixing $\sin^2\alpha \sim 0.01 $
fits the data better than the model without $\nu_s$ mixing (``null''
model) $\Delta\chi^2 < 0$. Also notice that for $\sin^2 \alpha \aprle
0.04$, $C$ and $\tau$ are below $3 \%$ and then they quickly increase
with $\alpha$ reaching $12 - 13 \%$ for $\sin^2 \alpha = 0.08$.

\begin{table}[h]
\caption{Results of the $\chi^2-$analysis of the IceCube zenith angle
distribution. Shown are $\chi^2_{\rm min}$ as well as the best fit
values of the normalization parameter $C$ and tilt $\tau$ for given
values of $\Delta m_{03}^2$ and $\sin^2\alpha$ in the mass-mixing
scheme. }
\begin{center}
\begin{tabular}{ccccc}
\hline\hline
$\Delta m_{03}^2$~(eV$^2$) & $\sin^2\alpha$ & $\chi^2_{\rm min}$ 
& $C$ & $\tau$
\\ \hline
    & 0.005 & 14.09 & 0.991 &  0.0175 \\
    & 0.01  & 15.29 & 0.997 &  0.0086 \\
0.5 & 0.02  & 16.50 & 1.008 & -0.0082 \\
    & 0.04  & 18.88 & 1.032 & -0.0394 \\
    & 0.08  & 31.73 & 1.085 & -0.1238 \\
\hline
    & 0.005 & 14.56 & 0.991 &  0.0217 \\
    & 0.01  & 15.33 & 0.997 &  0.0126 \\
1.0 & 0.02  & 16.97 & 1.010 & -0.0052 \\
    & 0.04  & 20.19 & 1.033 & -0.0347 \\
    & 0.08  & 39.41 & 1.092 & -0.1344 \\
\hline
    & 0.005 & 14.40 & 0.991 &  0.0247 \\
    & 0.01  & 14.45 & 0.996 &  0.0184 \\
2.0 & 0.02  & 16.11 & 1.008 &  0.0043 \\
    & 0.04  & 21.87 & 1.034 & -0.0323 \\
    & 0.08  & 43.29 & 1.094 & -0.1298 \\
\hline
    & 0.005 & 14.03 & 0.991 &  0.0246 \\
    & 0.01  & 14.92 & 0.996 &  0.0166 \\
3.0 & 0.02  & 15.84 & 1.008 &  0.0079 \\
    & 0.04  & 18.66 & 1.033 & -0.0217 \\
    & 0.08 & 41.98 & 1.098 & -0.1387 \\
\hline
IceCube & sim. & 14.16 & 0.982 &  0.04024 \\
\hline
\end{tabular}
\end{center}
\end{table}

\begin{figure}[ht]
\begin{center}
\includegraphics[width=15cm]{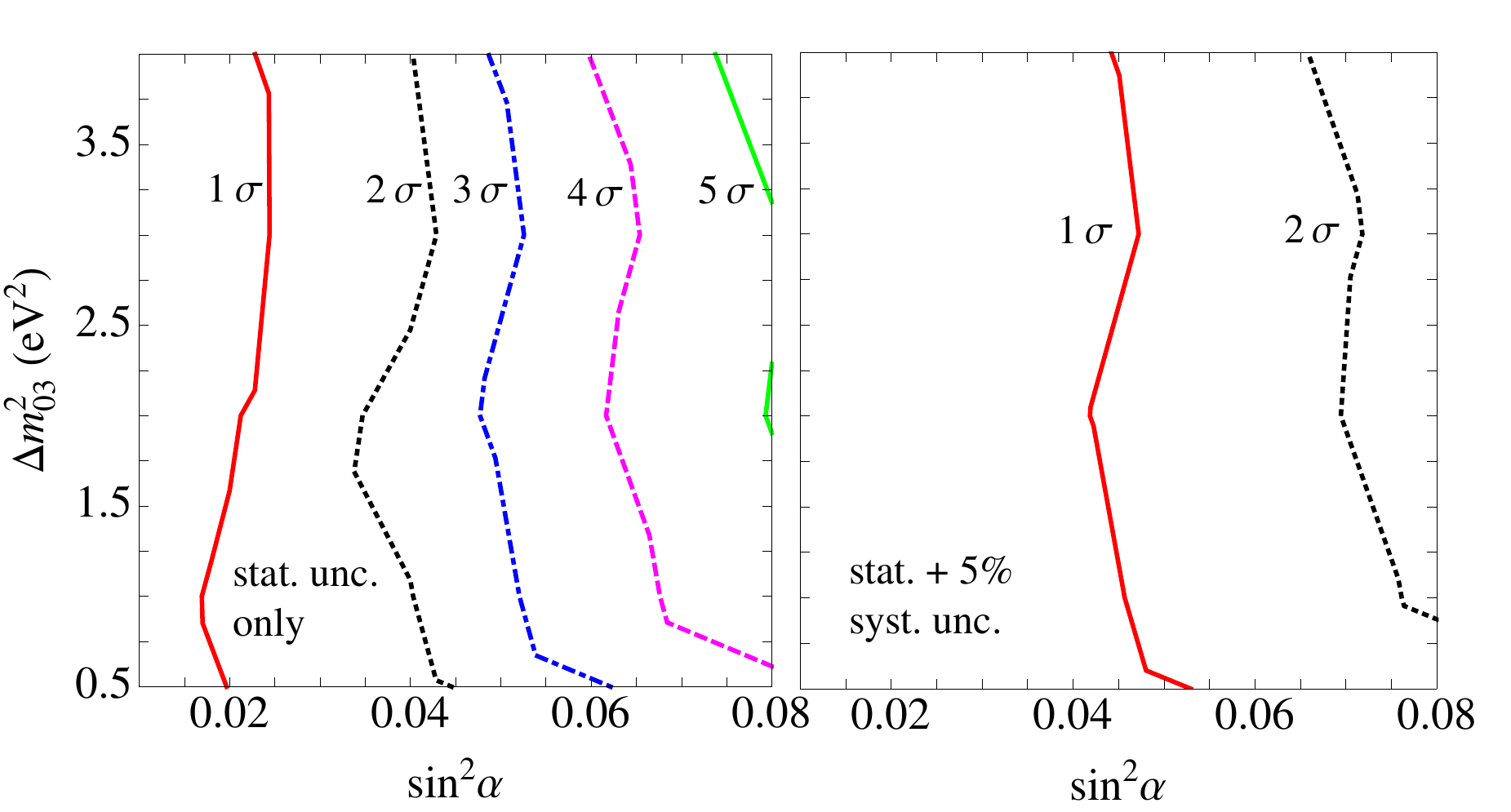}
\caption{Bounds on the active - sterile mixing angle as function of
$\Delta m_{03}^2$ obtained from the IceCube zenith angle distribution
of events.  Left panel - statistical errors only, right panel:
statistical plus $5\%$ uncorrelated systematic errors in each bin.}
\label{fig:fit}
\end{center}
\end{figure}

Fig.~\ref{fig:fit} (left panel) shows the bounds on the sterile
neutrino mixing as function of $\Delta m_{03}^2$ from the analysis
which takes into account statistical uncertainty in each bin as well
as the systematic uncertainties due to overal normalization and tilt
of the zenith angle distribution.  These are the main uncertainties.
To illustrate possible effect of other systematics we have taken the
extreme case: $5\%$ uncorrelated errors for individual bins (see
Fig.~\ref{fig:fit}, right panel). (Although it is expected that other
possible uncertainties are smooth functions of the zenith angle and
therefore correlate in different bins.)  In reality the effect of
additional errors should be smaller than that.  The parameter space to
the right hand side from the lines in Fig.~\ref{fig:fit} is excluded
at the indicated confidence level.

The bounds weakly depend on the $\Delta m_{03}^2$, as can be seen from
the behavior of the suppression factors (Figs.~\ref{fig:zen01} and
\ref{fig:zen1}).  The bounds are slightly weaker for smaller $\Delta
m_{03}^2$ since in this case the resonance dip shifts from the energy
range where IceCube has the highest sensitivity.

We find that with statistical uncertainties only (Fig.~\ref{fig:fit}
left panel) the upper bound is $\sin^2 \alpha < 0.05$ or $|U_{\mu
0}|^2 < 0.025$ at $3\sigma$ level and $\Delta m_{03}^2 = 1$ eV$^2$.
At $2\sigma$ level the bounds are $\sin^2 \alpha < 0.04$ and $|U_{\mu
0}|^2 < 0.02$.  At the same time interpretation of the LSND/MiniBooNE
results in terms of oscillations in the presence of sterile neutrinos
requires $|U_{\mu 0}|^2 \aprge 0.03$ for $\Delta m_{03}^2 = 1$ eV$^2$,
and $|U_{\mu 0}|^2 \aprge 0.06$ for $\Delta m_{03}^2 = 0.5$ eV$^2$.

With $5 \%$ uncorrelated systematic errors (Fig.~\ref{fig:fit} right
panel) the limits become substantially weaker: $\sin^2 \alpha = 0.06$,
is excluded at $90\%$ C.L.\ only.

\section{Oscillation effects in the  $\nu_s- \nu_\mu$  mixing scheme}

Let us consider the $\nu_s - \nu_\mu$ mixing only, {\it i.e.},
the simplest scheme of $\nu_s-$flavor mixing.  The corresponding
mixing matrix in the flavor basis $(\nu_s, \nu_\tau, \nu_\mu)$ equals
\be
{U}_f = U_{24} U_{23}  =
\left(\begin{array}{ccc}
c_{24} &  - s_{24} s_{23}  &   s_{24} c_{23}  \\
0 & c_{23} &   s_{23}   \\
- s_{24} & - c_{24} s_{23}  & c_{24} c_{23}
\end{array}
\right), 
\label{eq:f-mix}
\ee
where $s_{24} \equiv \sin \theta_{24}$, {\it etc.}.
Formally it differs from the mixing in (\ref{mass_scheme}) 
by permutation of $U_{23}$ with the $\nu_s-$mixing matrix.  

Now the mixing matrix elements, which determine the oscillations with
splitting $\Delta m_{03}^2$, equal $U_{s0} = c_{24}$, $U_{\tau 0} =
0$, $U_{\mu 0} = -s_{24}$.  They are reduced to the elements of our
simplest case (\ref{eq:u-elements}), if formally we take $c_{23} = 0$
and $s_{23} = -1$ and $\alpha = \theta_{24}$.  Therefore in the
leading order approximation for high energies the probabilities can be
obtained from the probabilities in $\nu_s-$mass mixing case by taking
$s_{23} = -1$. In particular, according to (\ref{eq:probcomp})
\be
P_{\mu \mu}^{(f)} \equiv |A_{\mu \mu}|^2 
\approx |A_{\tau^{\prime} \tau^{\prime}}(\theta_{24})|^2. 
\label{eq:prob-f}
\ee
For $A_{\tau^{\prime} \tau^{\prime}} = -1$ we obtain $P_{\mu \mu}^{f}
= 1$, whereas in the $\nu_s-$mass mixing scheme this value gives the
minimum of the dip $P_{\mu \mu}^{(mass)} = 0$.

It is possible to find relation between the sizes of dips for
different mixing schemes. For maximal 2-3 mixing we have from
(\ref{eq:probcomp}) the survival probability in the $\nu_s-$mass
mixing scheme (\ref{eq:mix-f}):
\be
P^{(mass)}_{\mu \mu} = \frac{1}{4} |A_{\tau^\prime \tau^\prime} + 1|^2. 
\label{eq:pmass}
\ee
In the resonance, the amplitude $A_{\tau^\prime \tau^\prime}$ is
approximately real.  This can be seen using explicit results for the
constant density case.  Indeed, according to (\ref{eq:h2}) in
resonance $H_{1m} = - H_{2m}$, and therefore (\ref{eq:ttamp}) gives
$A_{\tau^\prime \tau^\prime} \approx \cos (H_{1m}x)$.  Then from
(\ref{eq:pmass}) and (\ref{eq:prob-f}) we obtain relation between the
probabilities:
\be
P^{(f)}_{\mu \mu} = \left(2 \sqrt{P^{(mass)}_{\mu \mu}} - 1 \right)^2. 
\ee
Our numerical results in Fig.~\ref{fig:alter}  confirm this relation.

Let us consider corrections to the leading order result due to
oscillations driven by the 2-3 mixing and splitting.  They are
sub-dominant at high energies, but become dominant at low energies.
In the $\nu_s-$flavor mixing case it is convenient to consider
oscillations immediately in the flavor basis, {\it i.e.} take the
flavor basis as the propagation one.  Using the mixing matrix
(\ref{eq:f-mix}) we find the Hamiltonian of evolution $H = U_f
H^{diag} U_f^T + V $ which can be represented in the following form
\be
H =
\frac{\Delta m_{03}^2}{2 E}  
\left(\begin{array}{ccc}
c^2_{24} - \frac{2E V_{\mu}}{\Delta m_{03}^2} & 0  & 
- s_{24} c_{24}\\
0  & 0  & 0 \\
 - s_{24} c_{24}  & 0 & s^2_{24}  
\end{array}
\right) - 
\frac{\Delta m_{32}^2}{2 E}
\left(\begin{array}{ccc}
s^2_{24} c_{23}^2 & s_{24} s_{23} c_{23}  
&  s_{24} c_{24} c_{23}^2 \\
s_{24} s_{23} c_{23}   &  s_{23}^2  &  c_{24} s_{23} c_{23}  \\
s_{24} c_{24} c_{23}^2  &  c_{24} s_{23} c_{23} 
& c^2_{24} c_{23}^2
\end{array}
\right).  
\label{eq:halphaf}
\ee
At high energies the evolution is described by the first term of the
Hamiltonian (which does not depend on the 2-3 mixing), $\nu_\tau$
decouples and the corresponding $S$ matrix in the flavor basis can be
written as
\be
{S} =
\left(\begin{array}{ccc}
A_{ss} & 0  & A_{s \mu}   \\
0    &  1  &  0  \\
A_{\mu s}  & 0  &  A_{\mu \mu}
\end{array}
\right).
\label{eq:s-flav}
\ee
So that the survival probability, $P_{\mu \mu} = |A_{\mu \mu}|^2$, is
in accordance with (\ref{eq:prob-f}).  Indeed, the first term of the
Hamiltonian (\ref{eq:halphaf}) coincides with the Hamiltonian
(\ref{eq:halpha1}) up to permutation of the 2-3 lines, 2-3 columns and
substitution $\alpha \rightarrow \theta_{24}$, and therefore $A_{\mu
\mu} = A_{\tau^\prime \tau^\prime}$ in this approximation.  With the
sub-leading term of the Hamiltonian taken into account, evolution is
not reduced to the $2\nu-$evolution.

Effect of the 2-3 mixing at low energies ($E < 0.5$ TeV) can be
estimated in the following way.  In the basis $\nu_a$ defined in such
a way that $\nu_f = U_{24} \nu_a$ the Hamiltonian is given by
$$
H_a = U_{23} H^{diag} U_{23}^T + U_{24}^T V U_{24}~,
$$
or explicitly
\be
H_a =
\left(\begin{array}{lll}
\frac{\Delta m_{03}^2}{2 E}  - c^2_{24} V_\mu   & 0  & 
- s_{24}  c_{24}  V_\mu\\
0 & - s^2_{23}  \frac{\Delta m_{32}^2}{2 E}
&  -  s_{23} c_{23} \frac{\Delta m_{32}^2}{2 E}   \\
... &  ...  &   
- c_{23}^2  \frac{\Delta m_{32}^2}{2 E} - s_{24}^2 V_\mu
\end{array}
\right).
\label{eq:hgammalow}
\ee
For energies much below the sterile resonance, $V_\mu \ll \frac{\Delta
m_{03}^2}{2 E}$, one can perform a block diagonalization thus
decoupling the heaviest state, or simply neglect the 1-3 terms $s_{24}
c_{24} V_\mu$ in the Hamiltonian (\ref{eq:hgammalow}). The latter is
equivalent to an approximation of negligible matter effect on the
angle $\theta_{24}$. So, the evolution is reduced to $2\nu-$ problem.
Similarly to our consideration in Sec.\ 2 we find (returning to the
flavor basis) that the $\nu_\mu - \nu_\mu$ survival probability equals
\be
P_{\mu \mu} \approx |c_{24}^2 A_{\mu \mu}^{(a)} 
+ s_{24}^2 A_{s s}^{(a)}|^2,
\ee
where $A_{s s}^{(a)} = \exp[- ix (\frac{\Delta m_{03}^2}{2 E} -
c^2_{24} V_\mu)]$,  and the amplitude $A_{\mu \mu}^{(a)}$ should be
obtained by solving the evolution equation with the Hamiltonian
\be
H_a^{(2)} \approx
- \left(\begin{array}{ll}
0  &  \sin 2\theta_{23} \frac{\Delta m_{32}^2}{4 E}   \\
\sin 2\theta_{23} \frac{\Delta m_{32}^2}{4 E}  & 
\cos 2\theta_{23} \frac{\Delta m_{32}^2}{2 E} 
+ s_{24}^2 V_\mu
\end{array}
\right). 
\label{eq:hgammalow1}
\ee
Here we have subtracted from the $2 \times 2$ submatrix of
(\ref{eq:hgammalow}) the matrix proportional to the unit matrix.  The
$\nu_\mu - \nu_\mu$ probability averaged over fast oscillations driven
by $\Delta m_{03}^2$ equals
\be
P_{\mu \mu} = c_{24}^4 |A_{\mu \mu}^{(a)}|^2 + 
s_{24}^4 \approx  c_{24}^4 |A_{\mu \mu}^{(a)}|^2.
\ee

The matter effect on the amplitude $A_{\mu \mu}^{(a)}$ becomes
substantial when
$$
\frac{\Delta m_{32}^2}{4 E}  \sim  s_{24}^2 |V_\mu|,
$$
{\it i.e.}, $E \sim 150\,(0.04/ s_{24}^2)~ {\rm GeV}$.  Matter
suppresses the depth of $\nu_\mu - \nu_\tau$ oscillations and
increases the phase velocity as compared to the vacuum oscillation
case. For the maximal 2-3 mixing the effect is the same in the
neutrino and antineutrino channels:
$$
\phi_{32} = x  \sqrt{(\Delta m_{32}^2 / 2 E)^2 +
(s_{24}^2 V_\mu)^2}.  
$$
For non-maximal 2-3 mixing the resonance is realized at
$$
E = - \frac{\Delta m_{32}^2}{2 s_{24}^2 V_\mu } \cos 2\theta_{23},
$$
and the picture becomes $\nu - \bar{\nu}$ asymmetric depending on
$\cos 2\theta_{23}$.  We find that for $\theta_{23} = \pi/4$,
$s_{24}^2 = 0.04$ and $\cos \theta_z = - 1.0$ the averaged (over fast
oscillations) corrections to the probabilities in both channels equal
$\Delta P_{\mu \mu} \approx 0.15$ at $E = 100$ GeV and $\Delta P_{\mu
\mu} \approx 0.02$ at $E = 300$ GeV.

\begin{figure}[ht]
\begin{center}
\includegraphics[width=13cm]{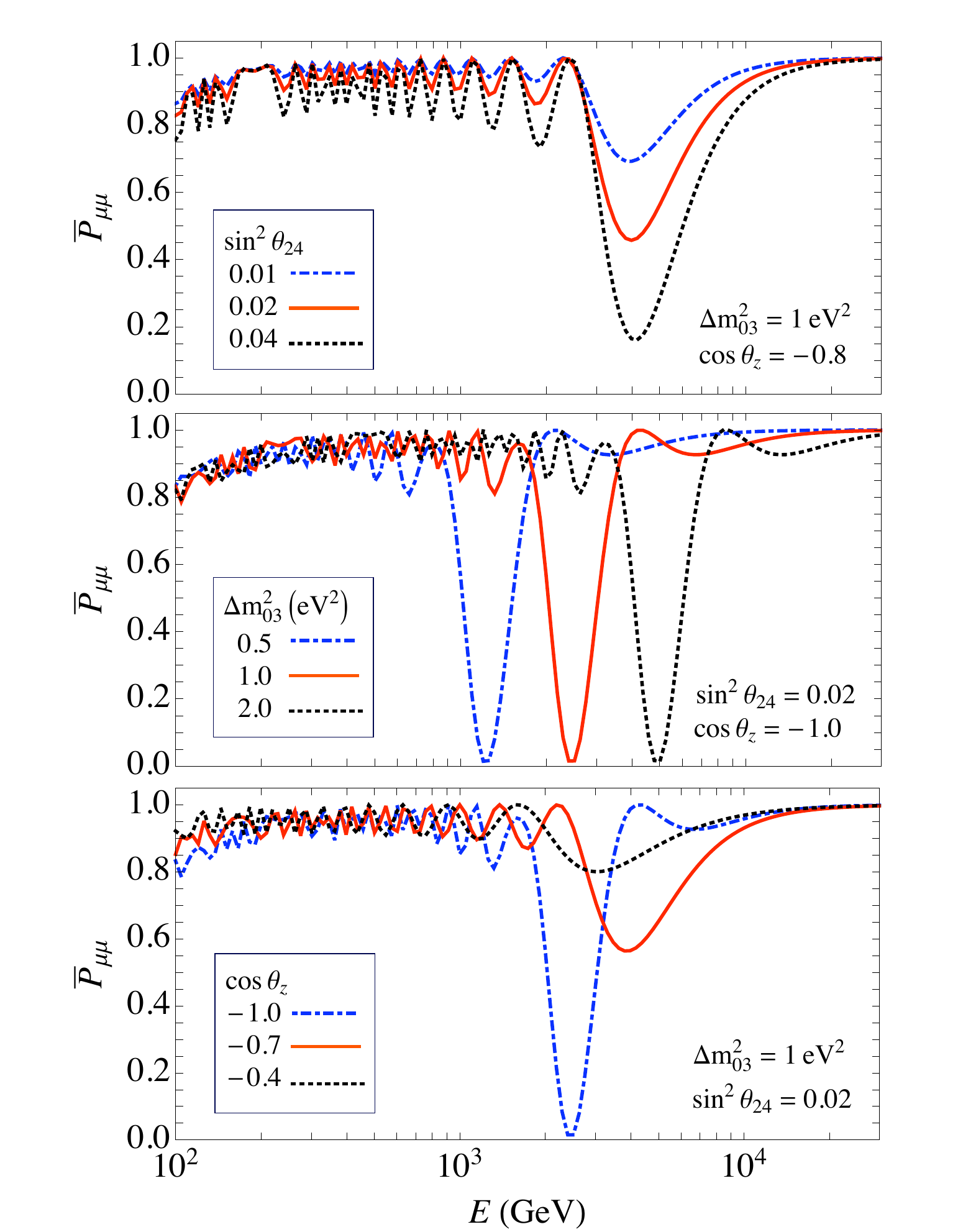}
\caption{The same as in Fig.~\ref{fig:probz1} for the $\nu_s -
\nu_\mu$ mixing scheme.  }
\label{fig:probz1f}
\end{center}
\end{figure}

\begin{figure}[ht] 
\begin{center} \includegraphics[width=13cm]{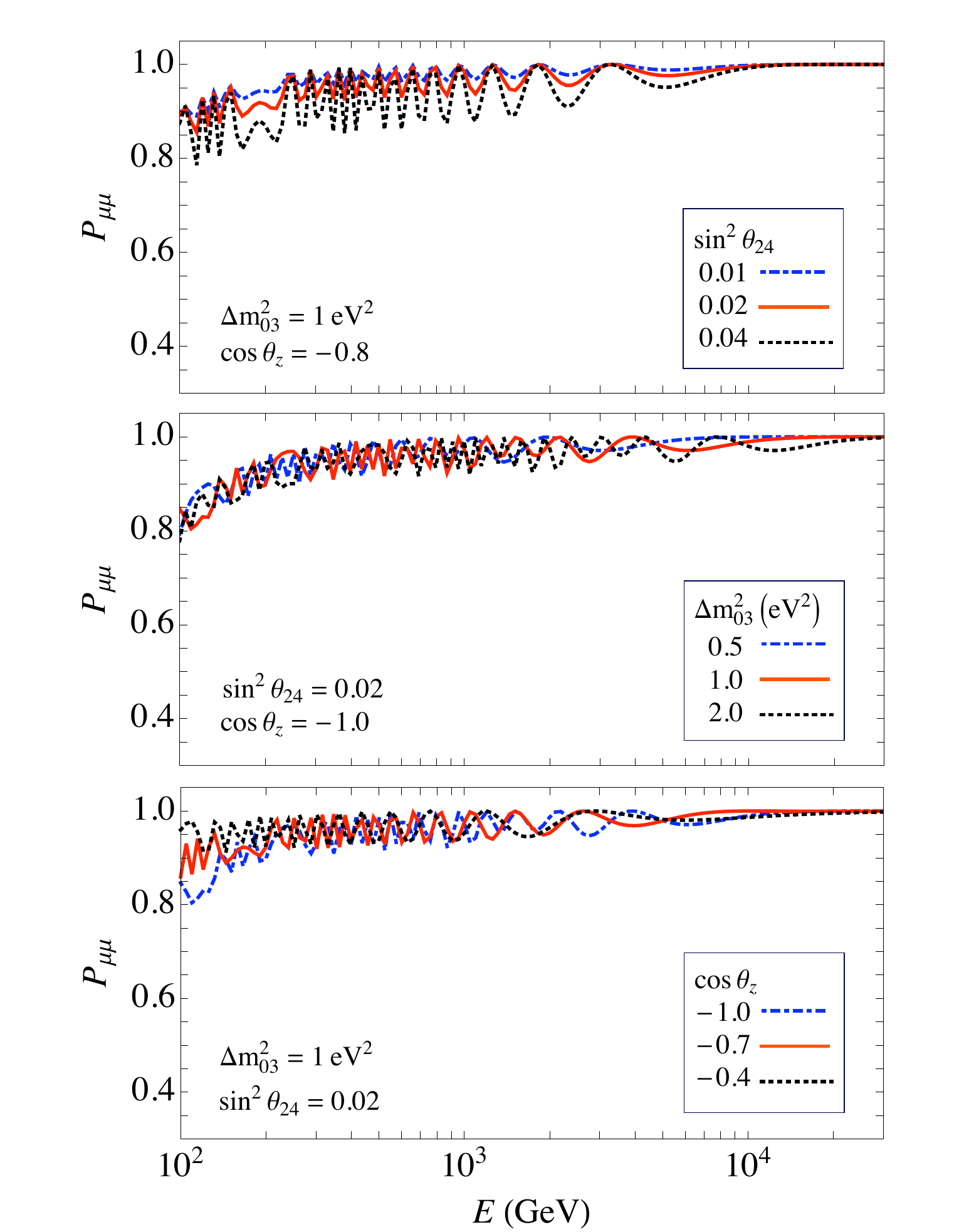} \caption{The same as
in Fig.~\ref{fig:probz2} for the $\nu_s - \nu_\mu$ mixing scheme.}
\label{fig:probz2f} 
\end{center} 
\end{figure} 

Results of numerical computations of the probabilities shown in
Figs.~\ref{fig:probz1f} and \ref{fig:probz2f} confirm this analytic
considerations.  Qualitatively the probabilities as functions of the
neutrino energy look rather similar to those in the $\nu_s-$mass
mixing scheme. As we discussed, certain difference appears at low
energies.  We find also that at $\sin^2 \theta_{24} = 0.08$ the dip
for $\cos \theta_z = -1$ is suppressed and maximal suppression is
achieved at $\cos \theta_z = - 0.90$, in contrast to the mass-mixing
case.  Also here the size of the dip decreases slower with increase of
$\cos \theta_z$.  This result holds for bigger mixing angles: If
$\sin^2 \theta_{24} = 0.08$, in the vertical bin we have $P_{\mu
\mu}^{(f)} (\cos \theta_z = - 1) \approx 1$, and maximal suppression
in the dip, $P_{\mu \mu}^{(f)} = 0$, is achieved at $\cos \theta_z = -
0.80$. Here in the dip region $\nu_\mu$ is transformed mainly to
$\nu_s$. So, the appearance of $\nu_\tau$ is the signature of the
$\nu_s-$mass mixing scheme.

\begin{figure}[ht]
\begin{center}
\includegraphics[width=11cm]{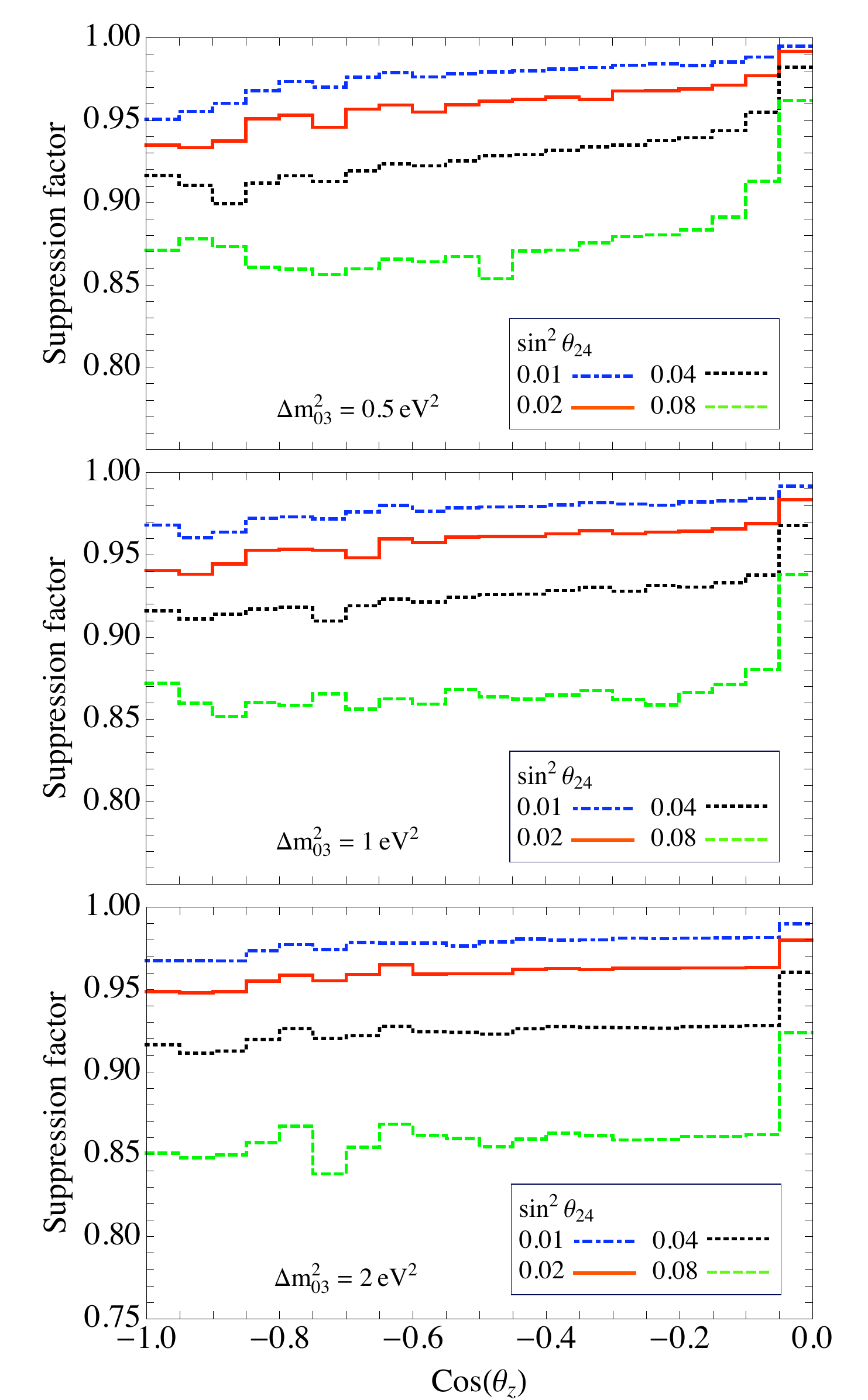}
\caption{The same as in Fig.~\ref{fig:zen01}
for the $\nu_s - \nu_\mu$ mixing scheme.}
\label{fig:zen01f}
\end{center}
\end{figure}

\begin{figure}[ht]
\begin{center}
\includegraphics[width=11cm]{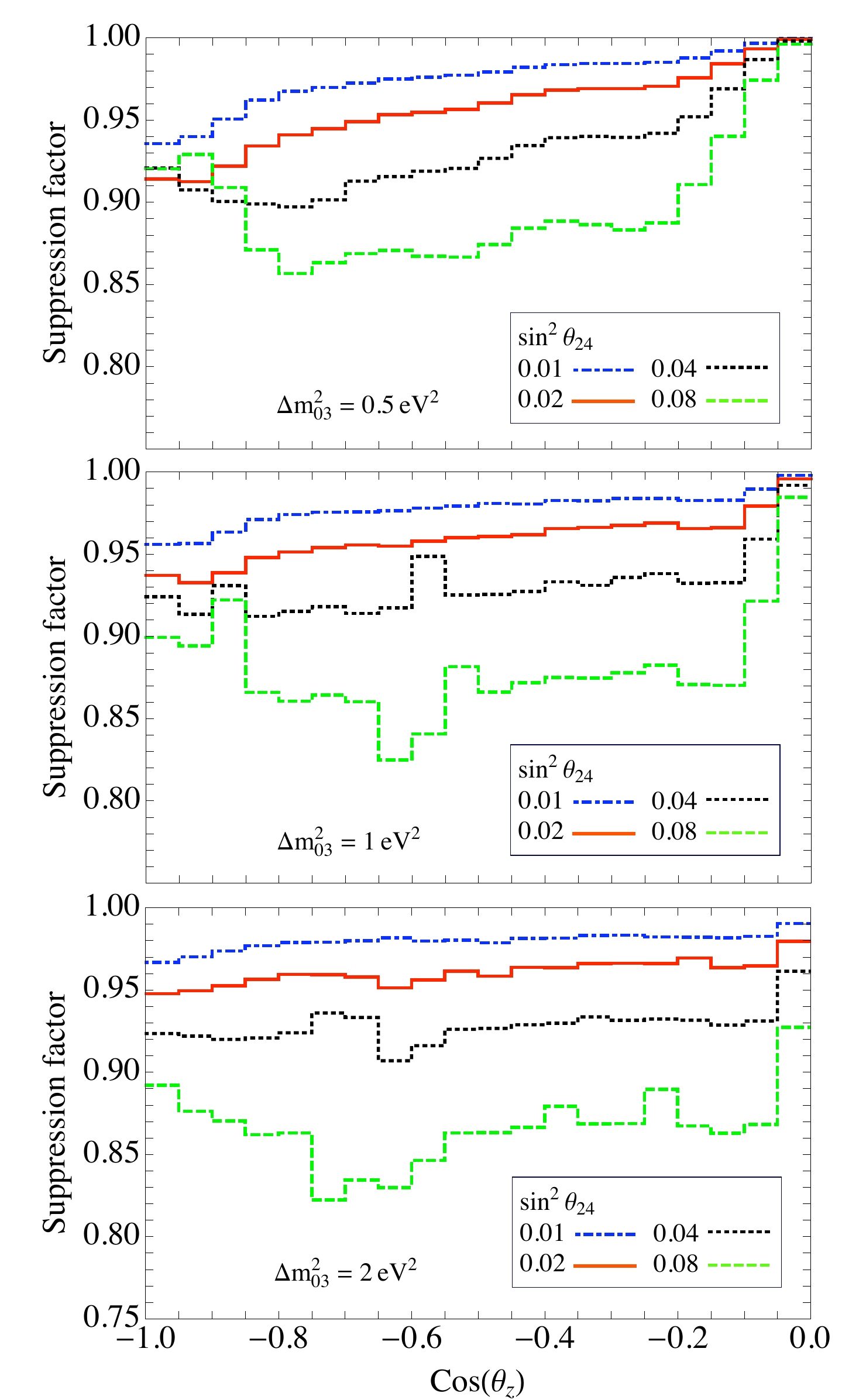}
\caption{The same as in fig. \ref{fig:zen1} for the $\nu_s - \nu_\mu$
mixing scheme.}
\label{fig:zen1f}
\end{center}
\end{figure}

In Figs.~\ref{fig:zen01f} and \ref{fig:zen1f} we present the zenith
angle dependence of the suppression factor for the muon events
integrated over the energy from $E_{th} =0.1$ TeV and $E_{th} =1$ TeV,
correspondingly. We compute these dependences in the same way as we
did for the $\nu_s-$mass mixing scheme.  Notice that for $\sin^2
\theta_{24} \leq 0.04$ the distributions are flatter than in
Figs.~\ref{fig:zen01} and \ref{fig:zen1}. The suppression is somewhat
stronger in vertical and nearly vertical bins and it is weaker in the
horizontal direction. In contrast to the previous scheme the
distribution changes with the threshold more strongly.  For $\sin^2
\theta_{24} = 0.08$, which is essentially excluded by MINOS result, a
wide dip appears in the range $\cos \theta_z = (-0.8, -0.4)$ (see
discussion in Sec.~6).

\begin{figure}[ht]
\begin{center}
\includegraphics[width=11cm]{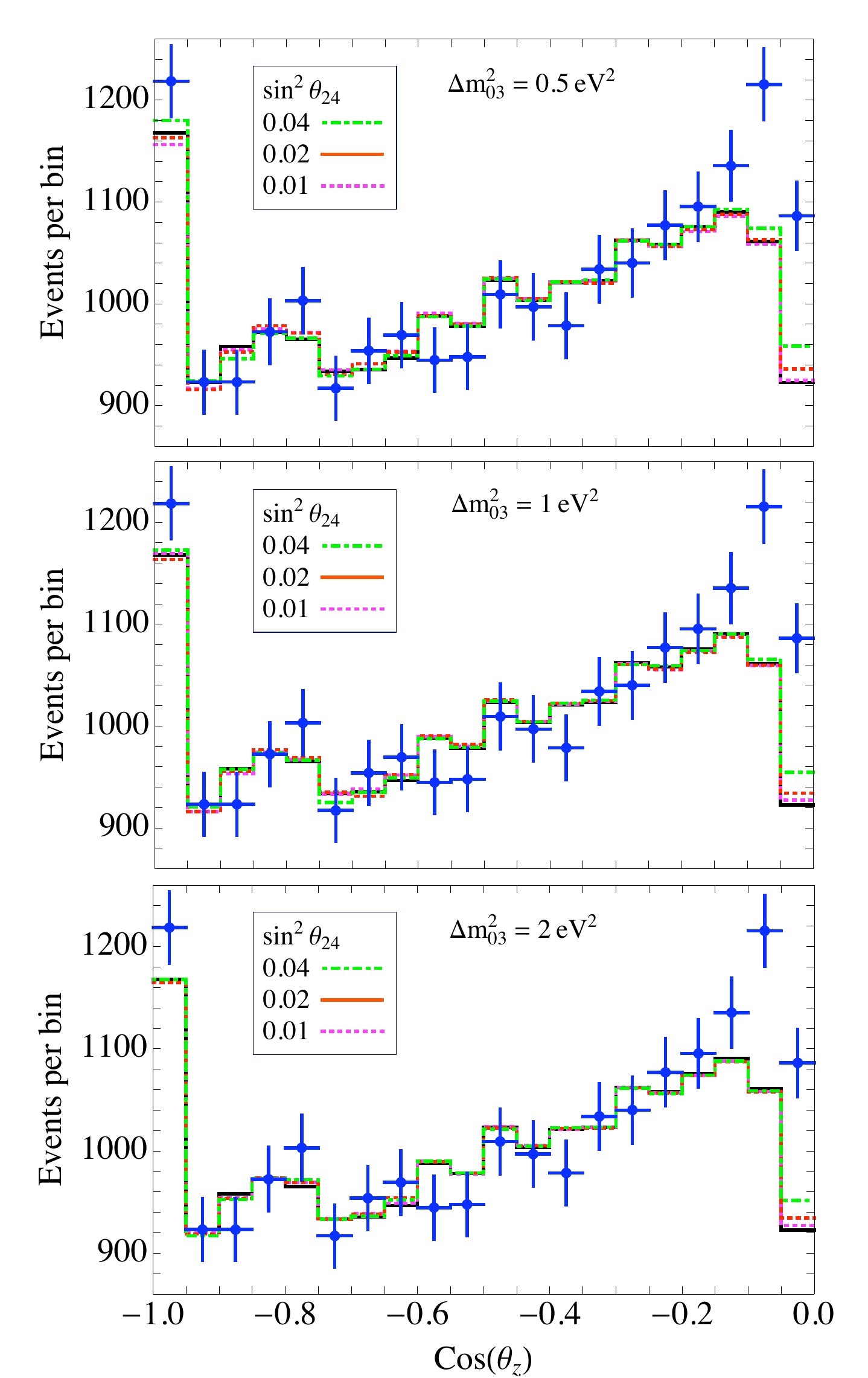}
\caption{The zenith angle distribution of muons from $\nu_\mu$
interactions integrated over the energy with oscillations to sterile
neutrinos in the $\nu_s - \nu_\mu$ mixing scheme.  We have
renormalized the event distribution according to the best-fit
normalization and tilt parameters from the $\chi^2$ fit (Table 2).
Also shown are the IceCube results.}
\label{fig:ice8f}
\end{center}
\end{figure}

In Fig.~\ref{fig:ice8f} we show the zenith angle distributions of the
$\mu-$events. The distributions are very similar (with some small
deviations in the vertical and horizontal bins) to those in the null
hypothesis case.

In Table~2 we present results of the $\chi^2$ analysis of the zenith
angle distribution for the $\nu_s - \nu_\mu$ mixing scheme.  In
contrast to the $\nu_s-$mass mixing case, now better fit than in null
hypothesis case can be achieved for values of $\sin^2 \theta_{24} =
0.02 - 0.04$ and $\Delta m_{42}^2 = (0.5 - 2)$ eV$^2$ which can
provide an explanation of the LSND/MiniBooNE results. So, $\nu_s$ with
these parameters can not be excluded by the present IceCube data.

\begin{table}[h]
\caption{Results of the $\chi^2-$analysis of the IceCube zenith angle
distribution. Shown are $\chi^2_{\rm min}$ as well as the best fit
values of the normalization parameter $C$ and tilt $\tau$ for given
values of $\Delta m_{03}^2$ and $\sin^2\theta_{24}$ in the $\nu_s -
\nu_\mu$-mixing scheme.}
\begin{center}
\begin{tabular}{ccccc}
\hline\hline
$\Delta m_{03}^2$~(eV$^2$) & $\sin^2\theta_{24}$ & $\chi^2_{\rm min}$
& $C$ & $\tau$
\\ \hline
    & 0.01  & 15.34 & 1.006 &  0.0052 \\
0.5 & 0.02  & 14.09 & 1.025 & -0.0023 \\
    & 0.04  & 11.92 & 1.060 & -0.0036 \\
    & 0.08  & 12.99 & 1.127 &  0.0176 \\
\hline
    & 0.01  & 13.93 & 1.005 &  0.0188 \\
1.0 & 0.02  & 15.20 & 1.025 &  0.0098 \\
    & 0.04  & 13.43 & 1.063 &  0.0137 \\
    & 0.08  & 12.80 & 1.138 &  0.0335 \\
\hline
    & 0.01  & 14.14 & 1.005 &  0.0240 \\
2.0 & 0.02  & 14.09 & 1.024 &  0.0227 \\
    & 0.04  & 13.68 & 1.063 &  0.0236 \\
    & 0.08  & 13.65 & 1.145 &  0.0256 \\
\hline
    & 0.01  & 15.11 & 1.005 &  0.0216 \\
3.0 & 0.02  & 14.43 & 1.024 &  0.0205 \\
    & 0.04  & 13.97 & 1.063 &  0.0271 \\
    & 0.08  & 19.67 & 1.149 &  0.0127 \\
\hline
IceCube & sim. & 14.16 & 0.982 &  0.04024 \\
\hline
\end{tabular}
\end{center}
\end{table}

\section{Oscillation effects for generic $\nu_s-$mixing in the leading approximation}

Let us consider the generic $\nu_s-$flavor mixing.  
The mixing matrix can be written as 
$U_f = U_{34} U_{24} U_{23}$, where $U_{34}$ is
the matrix of rotation in the $\nu_s - \nu_\tau$ plane on the angle
$\theta_{34}$. The matrix elements which describe the flavor content
of $\nu_0$ equal
\be
U_{s 0} = c_{34}  c_{24},~~ 
U_{\tau 0} = -s_{34} c_{24},~~ 
U_{\mu 0} = -s_{24}. 
\label{eq:usc}
\ee
The $\nu_\mu$ oscillations in vacuum (LSND/MiniBooNE) are determined
by the parameter $U_{\mu 0}$ -- the admixture of the muon neutrino in
the heaviest state.  In matter at high energies the phase $\phi_{32}$
is small and can be neglected, then the relevant parameters are
$U_{\mu 0}$, $U_{\tau 0}$, $U_{s 0}$. The Hamiltonian can be written
as
$$
H = \frac{\Delta m_{03}^2}{2 E} V_0 \times V_0^T + V
+ O\left( \frac{\Delta m_{03}^2}{2 E} \right),
$$
where $V_0^T \equiv (U_{s0}, ~ U_{\tau 0}, ~U_{\mu 0})$, and in the
first term we have the matrix formed by the product of the column
$V_0$ and the line $V^T_0$.

\begin{figure}[ht]
\begin{center}
\includegraphics[width=13cm]{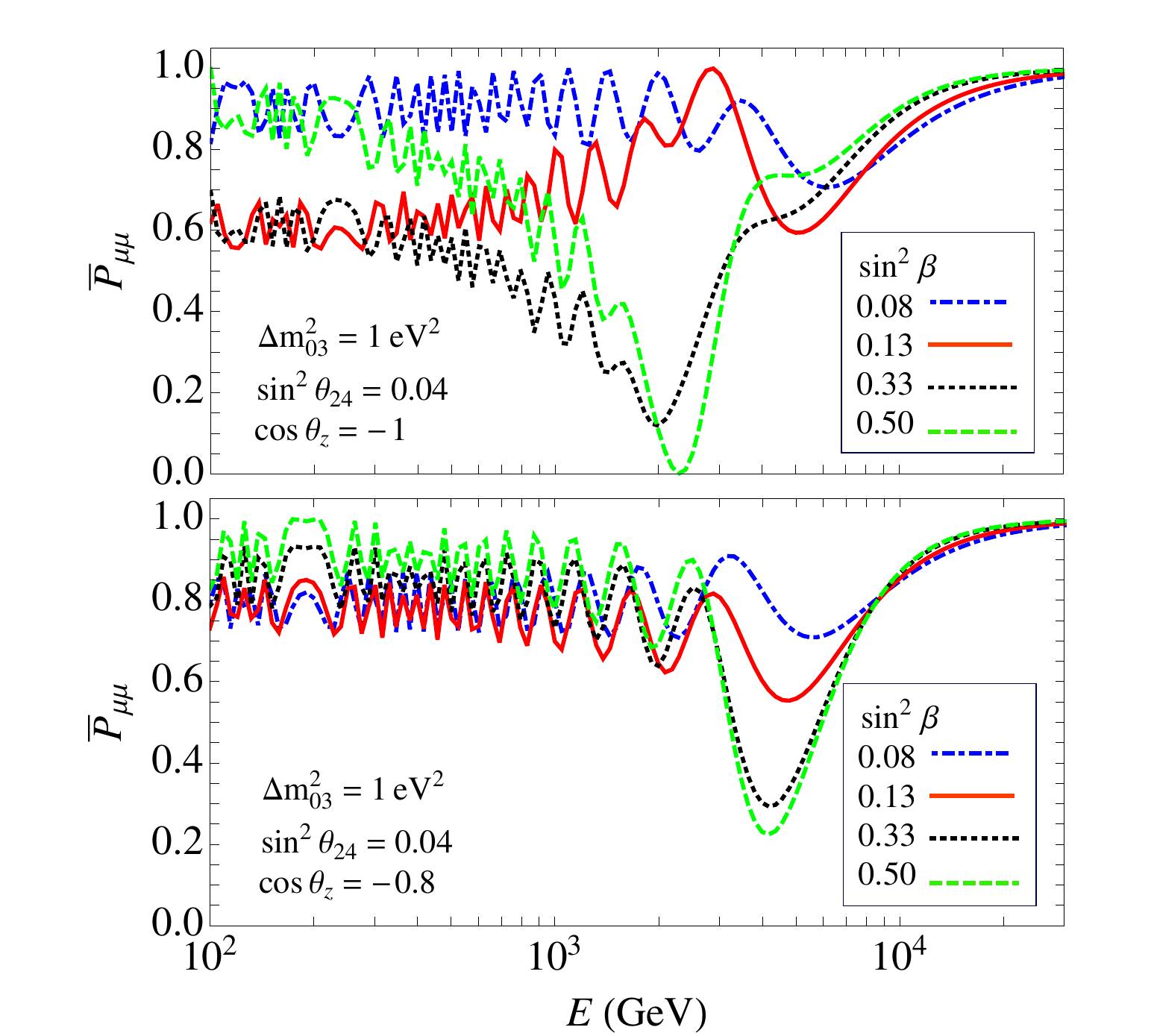}
\caption{The survival probabilities in the antineutrino (resonance)
channel as functions of neutrino energy for different mixing schemes
in the leading order approximation.  The {\em top and bottom panels}
show the probabilities for different values of $\sin^2\beta$
($s^2_{34}$) and two different zenith angles, while keeping $s_{24}^2$
fixed.  }
\label{fig:alter}
\end{center}
\end{figure}

Comparing (\ref{eq:usc}) with the elements in (\ref{eq:u-elements}) we
find that the dominant oscillation results at high energies in the
flavor case can be obtained from the results in the $\nu_s-$mass
mixing scheme identifying
$$
c_{34}  c_{24} = c_\alpha, ~~~ 
s_{24} = - s_\alpha s_{23}, ~~~
s_{34}  c_{24} = s_\alpha c_{23}. 
$$
That is, in general according to (\ref{eq:probcomp})
the probability equals 
\be
P_{\mu \mu} \approx 
\left|\sin^2 \beta ~A_{\tau^{\prime} \tau^{\prime}}(\alpha)
+ \cos^2 \beta \right|^2, 
\label{eq:pmmgen}
\ee
where $c_\alpha = c_{34} c_{24}$ or 
\be
s_\alpha^2 =  s_{24}^2 +  s_{34}^2 -   s_{34}^2 s_{24}^2 \approx  s_{34}^2 +  s_{24}^2, 
\label{eq:alphaeff}
\ee
and  
\be 
\sin^2 \beta
= \frac{s_{24}^2}{s_{24}^2 +  s_{34}^2 -   s_{34}^2 s_{24}^2} 
\approx \frac{s_{24}^2}{s_{24}^2 +  s_{34}^2}.
\label{eq:connect}
\ee
Explicitly,
\be
P_{\mu \mu} \approx  
\frac{1}{(1 - c_{24}^2  c_{34}^2)^2}
 \left|s_{24}^2 A_{\tau^{\prime} \tau^{\prime}}
+ s_{34}^2  c_{24}^2 \right|^2. 
\label{eq:probgen}
\ee

In Fig.~\ref{fig:alter} (top and bottom panels) we show the survival
probabilities as functions of energy for fixed value $s_{24}^2 = 0.04$
(as is required by LSND/MiniBooNE) and different values of
$\sin^2\beta$.  Notice that for the core crossing trajectories with
change of mixing scheme the size and form of the oscillation dip
changes significantly. The $\nu_s-$mass mixing case corresponds to
$\sin^2\beta = 0.5$ or $s_{24}^2 \approx s_{34}^2$, whereas the $\nu_s
- \nu_\mu$ mixing case is realized when $\sin^2\beta = 1$, that is
$s_{34}^2 = 0$.  Recall that at low energies the sub-leading effects
due to $\Delta m_{32}^2$ become important.

\begin{figure}[ht]
\begin{center}
\includegraphics[width=13cm]{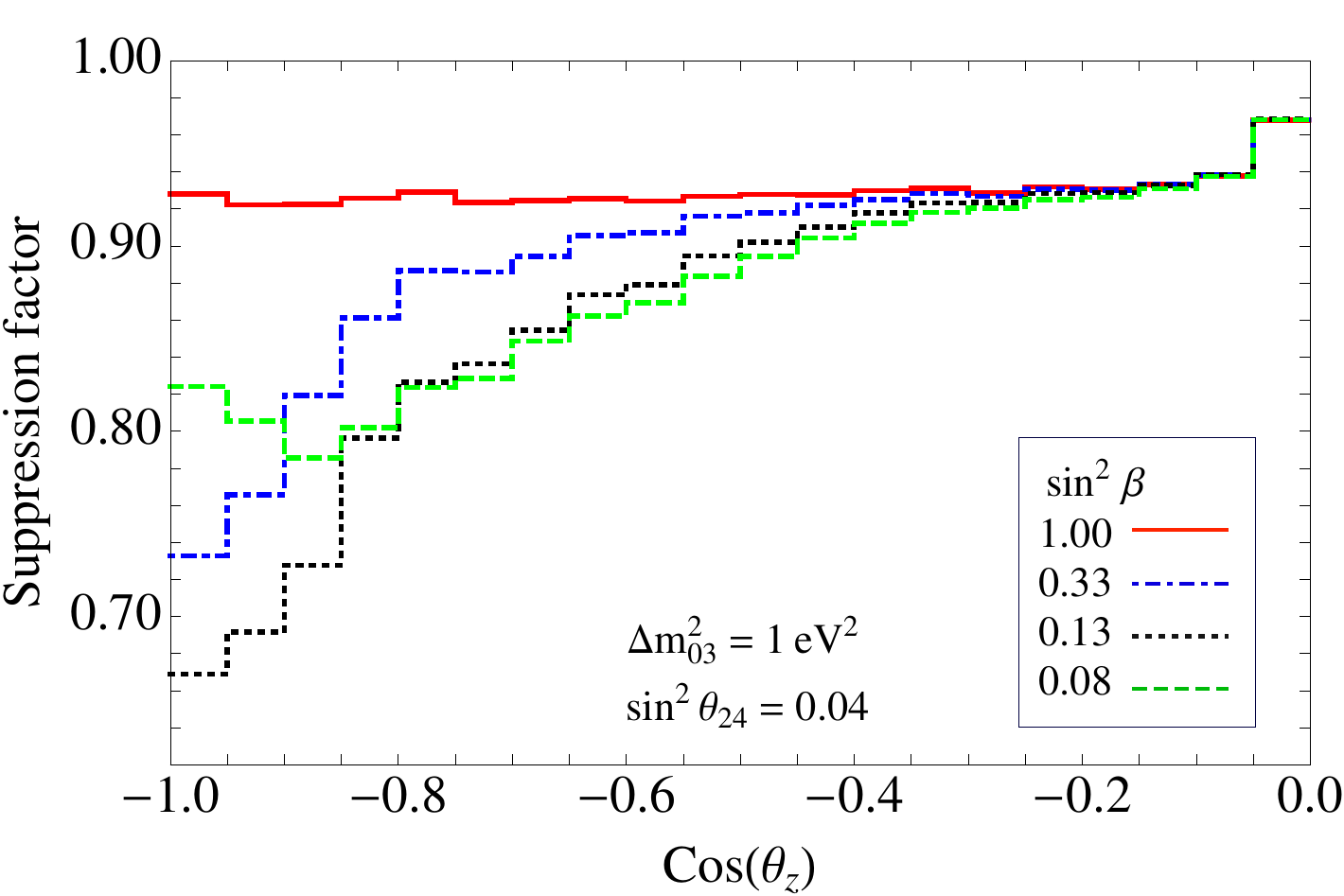}
\caption{The suppression factor in the leading order approximation as
function of the zenith angle for fixed $\theta_{24}$ and different
values of $\sin^2 \beta$.  We used $E_{th}=0.1$~TeV.  }
\label{fig:suppress2}
\end{center}
\end{figure}

In Fig.~\ref{fig:suppress2} we show the zenith angle dependence of the
suppression factor integrated over the energy (see definition in
(\ref{eq:indb})) for $s_{24}^2 = 0.04$ and different values of
$\sin^2\beta$.  Starting from $\sin^2\beta = 1$ ($s_{34}^2 = 0$) and
reducing it down to 0.08 one obtains first flat distribution, then the
distribution with a dip at or near the vertical directions and then
again rather flat distribution. In all the cases the suppression
weakens in the bins close to horizon.

The dip is at $|\cos \theta_z| \aprge 0.8$ in
Fig.~\ref{fig:suppress2}.  Indeed, in the $\nu_s - \nu_\mu$ mixing
case maximal suppression $P_{\mu\mu} = 0$ corresponds to
$A_{\tau^{\prime} \tau^{\prime}} = 0$. For the mantle-crossing
trajectories ($|\cos \theta_z| \aprle 0.8$) this can be achieved if
the MSW resonance condition and the oscillation phase condition
$\phi_{03} = \pi$ are satisfied simultaneously (see also discussion in
\cite{choubey}).  The conditions can be rewritten as
$$  
\frac{2\pi}{l_\nu} \cos 2\theta_{24} = V_\mu, 
~~~ 2x = \frac{l_\nu}{ \sin 2\theta_{24}},
$$
where ${l_\nu}$ is the oscillation length in vacuum, $x = 2R_E |\cos
\theta_z|$ is the length of neutrino trajectory ($R_E$ is the radius
of the Earth), and the expression in the RHS of the second equality
gives the oscillation length in resonance.  From these conditions,
excluding ${l_\nu}$, we find
\be
|\cos \theta_z| = \frac{\pi}{2 R_E V_\mu \tan 2 \theta_{24}} ~.
\label{eq:coscond}
\ee
Thus, a shift of the dip to small $|\cos \theta_z|$ would require
large $\nu_s - \nu_\mu$ mixing angle $\theta_{24}$.  The latter is
restricted by MINOS experiment \cite{minos}: $\sin^2 2 \theta_{24} <
0.14~~(90\% ~{\rm C.L.})$, and for the allowed values of $\theta_{24}$
the condition (\ref{eq:coscond}) can not be satisfied (see also
\cite{minos2}).  Large mixing $\alpha$ in the $2\nu$ amplitude
$A_{\tau^{\prime} \tau^{\prime}}$ is possible if $s_{34}^2$ is
large. However, in this case also $\sin^2\beta$ is substantially below
1. According to (\ref{eq:pmmgen}), $P_{\mu \mu} = 0$ corresponds to
$A_{\tau^{\prime} \tau^{\prime}}(\alpha) = - \cot^2 \beta$, {\it i.e.}
negative amplitude.  In turn, this requires even bigger phase than in
the previous case, $\phi_{03} > \pi$, which can not be achieved.

Notice that for values of oscillation parameters
\be
\Delta m_{03}^2 \sim (0.5 - 1)~ {\rm eV}^2,~~\sin^2 \alpha \sim 0.04, 
~~~\sin^2 \beta \sim 1 ,
\label{eq:special}
\ee 
the zenith angle distribution (suppression factor) for $|\cos
\theta_z| > 0.1$ is rather flat in spite of profound and wide dips in
the oscillation probabilities.  A shallow dip in the suppression
factor can appear in the interval of $\cos \theta_z$ $(-0.95, -0.8)$
for $\Delta m_{03}^2 \sim 0.5$ eV$^2$.  For $|\cos \theta_z| < 0.1$
the suppression becomes weaker which one can still use to disentangle
the oscillation effect and normalization of spectrum.  This flatness
of the energy integrated distribution is due to (i) specific
dependence of the IceCube sensitivity on energy and (ii) correlated
change of properties of the oscillation dip with change of $\theta_z$
which is realized for the parameters (\ref{eq:special}).

The zenith angle distribution with parameters (\ref{eq:special}) could
give even better fit, with a decrease in $\chi^2_{\rm min}$ value by
3, of the observed distribution than the null oscillation hypothesis.
Furthermore the required values of the overall normalization, 1.057,
and tilt, 0.0136, are small.  The contribution from low energy
oscillations driven by 2-3 mixing and mass splitting, however, has
strong dependence on the zenith angle, and consequently, distorts the
distribution near vertical directions.  Apparently study of the zenith
angle distributions with different energy threshoulds or in different
energy intervals will enhance sensitivity to oscillation effects.

Thus, apart from special case of $\nu_s-$flavor mixing in the leading
order approximation, the allowed mixing schemes predict the dip in the
zenith angle distribution in the vertical or nearly vertial
directions, and therefore are disfavored by the present IceCube data,
as in the illustrative analysis in Sec.~4.

Let us compare our results with those in Refs. \cite{nunokawa} and
\cite{choubey}.  In \cite{nunokawa} the flavor mixing has been
considered with $s_{24}^2 = 0.045$ and $s_{34}^2 = 0.45$ ({\it i.e.},
with nearly maximal $\nu_s - \nu_\tau$ mixing). According to
(\ref{eq:connect}) these parameters correspond to $\sin^2 \beta =
0.095$, and consequently, $P_{\mu \mu} \approx \left|0.095
A_{\tau^{\prime} \tau^{\prime}} + 0.905 \right|^2$.  This leads to
$\sim 10 - 20 \%$ effect with weak dependence on the zenith angle and
energy (see Fig.~4 in \cite{nunokawa}). Furthermore, since $s_\alpha^2
\approx 0.5$ (see (\ref{eq:alphaeff})), the mixing is nearly maximal
and therefore the resonance dip is absent (see Fig.~3e in
\cite{nunokawa}).  In \cite{choubey} the $\nu_s-$flavor mixing is
considered with $s_{24}^2 = s_{34}^2 = 0.04$. This equality means that
in fact the $\nu_s-$mass mixing is realized with $s_\alpha^2 \approx 2
s_{24}^2 = 0.08$.  Our results are in agreement with those in Fig.~5a
of \cite{choubey}.  Our interpretation of the dip at $\cos\theta_z =
-1$, however, differs: the dip is due to parametric enhancement of
oscillations, rather than the MSW oscillation dip in the medium with
averaged density.
 
Thus, the zenith angle and the energy distributions of events
substantially depend on details of the mixing scheme, and in
particular on mixing of $\nu_\tau$ in $\nu_0$ determined by $s_{34}$.

\section{Conclusions}

1. We have considered the neutrino oscillations in the Earth in the
presence of single sterile neutrino with mass $m \sim 1$ eV [$\Delta
m^2_{03} = (0.5 - 3)$ eV$^2$].

2. We present an analytic study of the oscillation probabilities which
allows one to understand features of dependences of the probabilities
on various parameters, and in particular, on the mixing scheme.  We
have identified the simplest mixing scheme in which the flavor
evolution is reduced to the $2\nu$ evolution.

3. The main features of probabilities (in the $\nu_s-$mass mixing
scheme) include the resonance dips (peaks) in the $\bar{\nu}-$channel
in the energy range $(0.5 - 5)$ TeV: the MSW resonance peak for the
mantle crossing trajectories and the parametric enhancement peak for
the core crossing trajectories.  In the $\nu-$channel at $E < 0.5$ TeV
there is the matter enhanced $\mu-\tau$ transition due to oscillations
induced by the 2-3 mixing and mass splitting.  The phase velocity in
the neutrino channel is enhanced due to matter effect, so that
oscillations are developed already for $E = (0.1 - 0.5)$ TeV. In
contrast, in the antineutrino channel, matter suppresses the phase
velocity and oscillations are not developed.

4. Oscillation effects on the $\nu_\mu-$ and $\bar{\nu}_\mu-$
atmospheric neutrino fluxes and on the sum of the fluxes are
studied. We have computed the energy spectra of these neutrinos,
integrated over various zenith angle intervals. Maximal oscillation
effect is $\sim 40\%$ suppression of the flux in a wide (half an order
of magnitute) energy interval.  The position of the dip is determined
by the value of $\Delta m^2_{03}$. The dip has low energy tail due to
oscillations driven by the 2-3 mixing. In the range (0.5 - 5) TeV the
effect is mainly due the resonance dip in the $\bar{\nu}-$channel,
whereas in the range (0.1 - 0.5) TeV it is mainly due to $\nu_\mu -
\nu_\tau$ oscillations with matter modified frequency in the
${\nu}-$channel.  Changes of the energy threshold does not modify
results substantially.

5. We have computed the zenith angle distributions of muon events
(induced by $\nu_\mu$ and $\bar{\nu}_\mu$) in the IceCube
detector. Oscillations lead to typical distortion of this distribution
with about $(1 - \sin^2 \alpha)$ suppression in the directions close
to the horizon, and stronger suppression in the directions close to
vertical $|\cos \theta_z| > 0.7$.  For the $\nu_s-$mass mixing scheme,
the maximal suppression, $20 - 25 \%$, is in the vertical direction.

6. The relative oscillation effect on the energy spectrum of neutrinos
can be enhanced by making integration over directions near the
vertical one.

7. We confronted the computed distributions with the IceCube data and
performed $\chi^2$ fit of the zenith angle distribution for the
$\nu_s-$mass mixing. We find that with statistical errors and
systematic uncertainties in the total normalization and tilt the
values, $|U_{\mu 0}|^2 > 0.025$ are excluded at more than $3\sigma$
level. The central value required by LSND/MiniBooNE is $|U_{\mu 0}|^2
\sim 0.03$ is excluded at the $3\sigma$ level.  With additional $5\%$
uncorrelated systematic uncertainties the limits become much weaker.

8. In the case of $\nu_s - \nu_\mu$ mixing scheme both properties of
the resonance dip and low energy behavior of the probabilities are
modified in comparison with those in the $\nu_s -$mass mixing scheme.
We find that maximal suppression is in the bins $\cos\theta_z = (-1.0,
-0.8)$.  The oscillation effects due to 2-3 mixing appears at $E <
0.3$ TeV both in neutrino and antineutrino channels, and the effects
are equal for maximal 2-3 mixing.  Rather flat zenith angle
distribution can be obtained in pure $\nu_\mu - \nu_s$ mixing case
with $|U_{\mu 0}|^2 \sim 0.04$ and $\Delta m_{03}^2 \sim 0.5$ eV$^2$
as well as for large $\nu_\tau - \nu_s$ mixing: $|U_{\tau 0}|^2 \sim
0.5$.  Fit to the zenith angle event distribution substantially
improves for this case and gives a better description of data than the
no $\nu_s-$ mixing case.

9. We have studied the oscillation effects in generic $\nu_s-$flavor
mixing scheme in the leading order approximations valid for high
energies $E \aprge 0.5$ TeV.  We showed how results for these schemes
can be obtained from the results of the $\nu_s-$mass mixing scheme.

10. Part of the parameter space of sterile neutrino ($U_{\mu 0}$,
$U_{\tau 0}$, $\Delta m_{03}^2$) relevant for the LSND/MiniBooNE can
be excluded by the the present IceCube 40 data. Namely, the region of
$|U_{\mu 0}| \sim |U_{\tau 0}| > 0.15$ and $\Delta m^2 > 0.8$ eV$^2$ is
excluded at about $3 \sigma$ level. At the same time in certain
regions of this parameter space, e.g. $|U_{\tau 0}| \sim 0$, $|U_{\mu
0}| = 0.13 - 0.27$ (which correspods to the $\nu_\mu$ flavor mixing)
one can obtain even better fit of the data than in the no
$\nu_s-$mixing case. Our analysis has an illustrative character and
complete scan of the whole parameter space is beyond the scope of this
paper.  Such an analysis can be done after release of new IceCube data
and better understanding of systematic errors. Substantial improvement
of sensitivity to sterile neutrino oscillations will be possible when
the two dimensional (energy-zenith angle) distribution of events will
be available~\cite{soeb2}.
That is, future studies of the zenith angle distributions with
different energy thresholds or in different energy intervals will
allow to perform very sensitive search for sterile neutrinos.

\section*{Acknowledgments}

We thank Doug Cowen, Kara Hoffman, Paolo Desiati, Elisa Resconi and
specially Warren Huelsnitz for helping us understand the IceCube
results better.  Work of S.R. was funded while under contract with the
U.S.~Naval Research Laboratory.

\section*{Appendix. Constant density case}

To a good approximation the case of constant density can be applied
for neutrinos crossing the mantle of the Earth only. For constant
$V_\mu$ the Hamiltonian (\ref{eq:halpha1}) can be diagonalized by the
rotation on the mixing angle in matter $\alpha_m$:
\be
\sin^2 2\alpha_m = \frac{\sin^2 2\alpha}{\left(\cos 2\alpha - 
\frac{2V_\mu E}{\Delta m_{03}^2}\right)^2 + \sin^2 2\alpha}.  
\label{eq:alphamix}
\ee
Integration of the evolution equation is then trivial, giving the
$S-$matrix
\be
S_m = 
\left(\begin{array}{ccc}
e^{-iH_{1m}x}  &  0   & 0  \\
0 &  e^{- i H_{2m} x}   & 0 \\
0 & 0 & e^{i \phi_{32}}
\end{array}
\right) ,     
\label{eq:diagmat}
\ee
where $H_{im}$ are the eigenvalues of the Hamiltonian in matter:  
\be
H_{1m, 2m} = \frac{1}{2} \left(\frac{\Delta m_{03}^2}{2E} - V_\mu \right) 
\pm \frac{\Delta m_{03}^2}{4E} 
\sqrt{\left(\cos 2\alpha - \frac{2E V_\mu}{\Delta m_{03}^2} \right)^2 
+ \sin^2 2\alpha}~.  
\ee
$H_{1m}$ corresponds to the $+$ sign. 
In the antineutrino (resonance) channel the eigenvalues,  
as functions of neutrino energy,  have  the following asymptotics: 
\be
H_{1m} \approx 
\left\{
\begin{array}{ll}
\frac{\Delta m_{03}^2}{2E} - V_\mu \cos^2 \alpha, & E \rightarrow 0; \\ 
\frac{\Delta m_{03}^2}{4E} \sin 2\alpha, & {\rm resonance};\\
\frac{\Delta m_{03}^2}{2E} \sin^2 \alpha, & \frac{\Delta m_{03}^2}{2E}  
\ll  V_\mu. 
\end{array}
\right.
\label{eq:h1}
\ee
\be
H_{2m} \approx 
\left\{
\begin{array}{ll}
- V_\mu \sin^2 \alpha,~~~~E \rightarrow 0; \\
- \frac{\Delta m_{03}^2}{4E} \sin 2\alpha, & {\rm resonance}; \\
- V_\mu & \frac{\Delta m_{03}^2}{2E}  \ll V_\mu. 
\end{array}
\right.
\label{eq:h2}
\ee
Since for antineutrinos $V_\mu > 0$, one has $H_{2m} < 0$.  In the limit of high energies: 
$H_{1m} - H_{2m} = V_\mu$.

Returning back to the $\tilde{\nu}$  basis,  
$\tilde{S} =  U(\alpha_m) S_m  U^{\dagger}(\alpha_m)$,  we obtain  
\be 
A_{\tau^{\prime} \tau^{\prime}} = \sin^2 \alpha_m  e^{-iH_{1m}x}
+ \cos^2 \alpha_m  e^{-iH_{2m}x}. 
\label{eq:ttamp}
\ee  
Then insertion of this amplitude in (\ref{eq:probcomp}) gives 
\be
P_{\mu \mu}  =  \left|\sin^2 \theta_{23}\left(\sin^2 \alpha_m  e^{-iH_{1m}x}
+ \cos^2 \alpha_m  e^{-iH_{2m}x} \right) 
+ \cos^2 \theta_{23}  e^{i\phi_{32}} \right|^2, 
\ee
and explicitly: 
\be
P_{\mu \mu}  =  1 - \sin^4 \theta_{23} 
\sin^2 2\alpha_m \sin^2 (\phi_1 - \phi_2) - 
\sin^2 2\theta_{23} (\sin^2\alpha_m  \sin^2 \phi_1 + \cos^2\alpha_m  \sin^2 \phi_2). 
\label{eq:mumuconst}
\ee
Here 
\bea
\phi_1 & = & \frac{1}{2}(H_{1m} x + \phi_{32}) = \frac{1}{2}
\left(H_{1m} + \frac{\Delta m^2_{32}}{2E} \right)x, 
\nonumber\\
\phi_2 & = & \frac{1}{2}(H_{2m} x + \phi_{32}) =  
\frac{1}{2}\left(H_{2m} +  
\frac{\Delta m^2_{32}}{2E} \right)x,
\label{eq:phases12}
\eea
and consequently,  
\be
\phi_1 - \phi_2 = \frac{1}{2}(H_{1m} - H_{2m}) x.  
\ee

If $E \rightarrow 0$ (vacuum oscillation limit), $H_{2m} \rightarrow
0$ and $H_{1m} \rightarrow \frac{\Delta m^2_{03}}{2E}$.  Therefore
\be
\phi_1 \rightarrow  \frac{\Delta m^2_{02}}{4E} x, ~~~
\phi_2 \rightarrow \frac{\Delta m^2_{32}}{4E} x, ~~~
\phi_1 - \phi_2 = \frac{\Delta m^2_{03}}{4E} x. 
\ee
In this case also $\alpha_m \rightarrow \alpha $ and the averaged over
fast oscillations probability equals
\be
\bar{P}_{\mu \mu}  =  1  
-  \cos^2\alpha \sin^2 2\theta_{23} \sin^2 \frac{\Delta m^2_{32} x}{4E} 
- 0.5 \sin^4 \theta_{23} \sin^2 2\alpha 
- 0.5 \sin^2 2\theta_{23} \sin^2\alpha.  
\label{eq:mumuconst1}
\ee
The first two terms correspond (up to $\cos^2\alpha$) to the standard
2-3 probability and corrections are of the order $\sin^2\alpha$.

In the limit of high energies for antineutrinos we have 
$H_{1m} \rightarrow \frac{\Delta m_{03}^2}{2E} \sin^2\alpha$ 
and  $H_{2m} \rightarrow - |V_\mu|$. 
So,  
\be
\phi_1  = \frac{1}{2} \left(\frac{\Delta m_{03}^2}{2E} \sin^2\alpha 
+ \frac{\Delta m^2_{32}}{2E} \right)x 
\approx \frac{\Delta m_{03}^2 x}{4E} \sin^2\alpha,
\ee
\be
\phi_2 = \frac{1}{2}(- |V_\mu| x + \phi_{32}) 
\approx - \frac{1}{2}|V_\mu| x. 
\ee
For high energies (above the resonance): $\alpha_m \approx 90^{\circ}$,  
and consequently,  
\be
P_{\mu \mu}  \approx  
1 - \sin^2 2\theta_{23} \sin^2 \phi_1 \approx  1 - \sin^2 2\theta_{23} 
\sin^2 \left(\frac{\Delta m_{03}^2 x}{4E} \sin^2\alpha \right). 
\label{eq:mumuconst2}
\ee
In the case of constant density 
the  $\nu_{\tau}^{\prime} - \nu_\tau^{\prime}$ probability 
is described by   usual oscillation formula: 
\be
P_{\tau^{\prime} \tau^{\prime}} = 
1- \sin^2 2\alpha_m \sin^2 \phi_m , 
\ee
where $\phi_m$  is the half-phase of oscillations in matter:  
\be
\phi_m = \frac{\Delta m_{03}^2 x}{4E} \sqrt{\left(\cos 2\alpha -
\frac{2V_\mu E}{\Delta m_{03}^2}\right)^2 + \sin^2 2\alpha}.  
\ee


\end{document}